\documentclass[twocolumn,aps,prl,amsmath,superscriptaddress,notitlepage,longbibliography,nobibnotes]{revtex4-2}
\usepackage{amssymb}
\usepackage{amsmath}
\usepackage{mathrsfs}
\usepackage{appendix}
\usepackage{graphicx}
\usepackage{float}
\usepackage[colorlinks=True,citecolor=blue,linkcolor=blue,allcolors=blue]{hyperref}
\usepackage[normalem]{ulem}
\usepackage{color}
\usepackage{bm}

\newcommand{\GJ}{\textcolor{black}}

\newcommand{\WW}{\textcolor{black}}

\begin{document}
	
	\title{Lateral Shift as a Control Knob for Localization Transitions in a Quasiperiodic Ladder}
	
	\author{Bing Shao}
	\thanks{These authors have contributed equally to this work.}
	\affiliation{College of Physics and Optoelectronic Engineering, Ocean University of China, Qingdao 266100, China}
	
	\author{Guangjie Zhang}
	\thanks{These authors have contributed equally to this work.}
	\affiliation{College of Physics and Optoelectronic Engineering, Ocean University of China, Qingdao 266100, China}
	
	\author{Longwen Zhou}
	\email{zhoulw13@u.nus.edu}
	\affiliation{College of Physics and Optoelectronic Engineering, Ocean University of China, Qingdao 266100, China}
	\affiliation{Qingdao Key Laboratory of Advanced Optoelectronics, Qingdao 266100, China}
	\affiliation{Engineering Research Center of Advanced Marine Physical Instruments and Equipment of MOE, Qingdao 266100, China}
	
	\author{Jiangbin Gong}
	\email{phygj@nus.edu.sg}
	\affiliation{Department of Physics, National University of Singapore, 117551, Singapore}
	\affiliation{Centre for Quantum Technologies, National University of Singapore, 117543, Singapore}
	\affiliation{MajuLab, CNRS-UCA-SU-NUS-NTU International Joint Research Unit, Singapore.}
	
	\author{Weiwei Zhu}
	\email{phyzhuw@tongji.edu.cn}
	\affiliation{College of Physics and Optoelectronic Engineering, Ocean University of China, Qingdao 266100, China}
	\affiliation{Center for Phononics and Thermal Energy Science, China-EU Joint Lab on Nanophononics, Shanghai Key Laboratory of Special Artificial Microstructure Materials and Technology, School of Physics Science and Engineering, Tongji University, Shanghai 200092, China}
	
	\begin{abstract}
		\GJ{This work reports rich localization-delocalization transitions in a quasiperiodic ladder, of which the two legs are subject to the same quasiperiodic onsite potential but can be shifted laterally relative to each other.  It is found that the lateral shift between the two legs  effectively generates a magnetic flux in the reciprocal momentum space.  The lateral shift thus offers a control knob, allowing us to access and simulate rich phenomena including magnetic-flux-enhanced localization, magnetic-flux-suppressed localization, and magnetic-flux-induced reentrant localization transitions. The underlying physical mechanisms  as well as the phase boundaries separating localized, mixed, and extended phases are both qualitatively and quantitatively understood, \WW{based on a band-structure analysis that employs a commensurate approximation to the quasiperiodic potential, requiring only unit cells of small to modest sizes.} Our work provides a highly tunable platform for exploring localization physics with promising applications such as quantum switching, and a {broadly applicable approach} for understanding {localization-delocalization} transitions in quasiperiodic systems.}
	\end{abstract}
	
	\maketitle
	
	\textit{Introduction.---}Anderson localization describes how disorder may halt wave propagation and induce exponential localization~\cite{PhysRev.109.1492,RevModPhys.80.1355}.
	Specifically, it manifests a sharp dimensional dependence: while a finite disorder strength is needed to drive the Anderson transition in three dimensions, 
	In one-dimensional (1D) and two-dimensional (2D) disordered systems, any infinitesimal disorder suffices to localize all eigenstates~\cite{PhysRevLett.42.673}.  Only in three dimensions or higher can there be localization-delocalization transitions in disordered systems.
	Lying between clean and disordered systems, quasiperiodic systems provide an ideal controllable platform for investigating possible localization-delocalization transitions in 1D  systems~\cite{PhysRevLett.61.2144,PhysRevA.75.063404,PhysRevA.80.021603,PhysRevA.90.061602,PhysRevB.91.014108,PhysRevE.93.022209,PhysRevLett.120.160404,PhysRevLett.122.237601,PhysRevB.100.054301,PhysRevB.101.174205,PhysRevResearch.2.033052,Goblot2020,PhysRevB.103.054203,PhysRevLett.129.103401,PhysRevLett.129.113601,PhysRevB.105.174206,Weidemann2022,PhysRevLett.134.053601}, with the Aubry--Andr\'e (AA) model serving as a paradigmatic example~\cite{aubry1980analyticity}. This model exhibits self-duality at a critical quasiperiodic potential strength, separating a purely extended phase from a purely localized phase~\cite{rspa.1984.0016,A.P.Siebesma_1987,PhysRevA.36.5349,69337d76b53b,PhysRevLett.51.1198}. By breaking this self‑duality through modified couplings~\cite{PhysRevB.96.054202,PhysRevLett.123.025301,PhysRevB.100.174201,PhysRevB.106.024204,PhysRevB.105.014207,PhysRevB.107.075128,PhysRevB.98.104201} or on‑site potential modulations~\cite{PhysRevLett.60.1334,PhysRevLett.61.2141,PhysRevB.41.5544,PhysRevB.96.085119,PhysRevLett.123.070405,2rfb-j778}, generalized AA models have been constructed to study the physics 
	of mixed phases with mobility edges~\cite{PhysRevLett.104.070601,PhysRevLett.114.146601,PhysRevLett.125.196604,PhysRevLett.131.186303,PhysRevB.108.L100201,rl1f-ptzq,Li2025,PhysRevLett.126.106803,PhysRevB.105.L220201,PhysRevB.105.054204,PhysRevB.107.224201,PhysRevResearch.5.033170}.
	
	The physics and engineering of mixed phases with mobility edges in quasiperiodic ladders have attracted sustained interest for a long time~\cite{PhysRevLett.101.076803,flach2014,PhysRevLett.113.236403,PhysRevLett.116.140401,PhysRevX.8.031045,PhysRevB.99.054211}.  For ladder systems, studies to date have indicated that placing identical quasiperiodic onsite potentials on the two legs of a ladder cannot yield mixed phases and mobility edges~\cite{PhysRevLett.101.076803,flach2014,PhysRevB.99.054211}, a fact explainable in terms of decoupling such ladder systems into two independent AA models. Mixed phases and mobility edges can instead be achieved by introducing an antisymmetric quasiperiodic onsite potential in flat-band ladders~\cite{PhysRevLett.113.236403} or by applying a magnetic flux in a zigzag model~\cite{PhysRevX.8.031045}. Pursuing a different route, we discover in this work that a simple lateral shift between two quasiperiodic chains already suffices to yield rich mixed phases. The use of a lateral shift between two lattices echoes with ongoing wide research activities on bilayer moir\'{e} physics~\cite{PhysRevLett.99.256802,Trambly2010,pnas.1108174108}.

	\begin{figure}
		\includegraphics[width=\linewidth]{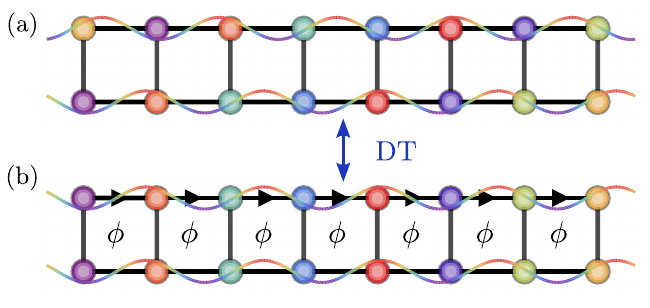}
		\caption{Quasiperiodic ladders. (a) A quasiperiodic ladder with a lateral shift between its upper and lower legs. (b) The duality-transformed (DT) quasiperiodic ladder with an synthetic magnetic flux $\phi$ per plaquette. With the rational approximation $F_5/F_6=5/8$ to the irrational $\alpha$ parameter, the lattice becomes periodic with a unit cell of 8 sites.}
		\label{model}
	\end{figure}
	
In this Letter, we report how the localization-delocalization transitions in 1D quasiperiodic lattices may be controlled by a lateral shift between two otherwise identical (upper and lower) legs, as illustrated in Fig.~\ref{model}(a).  Through a duality transformation, the lateral shift can be mapped to a Peierls phase in the intraleg coupling, effectively acting as a synthetic magnetic flux. It is this effective flux that explains why the lateral shift between two legs can act as an appealing control knob, leading to magnetic‑flux‑enhanced localization, magnetic‑flux‑suppressed localization, and magnetic‑flux‑induced reentrant localization transitions~\cite{PhysRevLett.126.106803,PhysRevB.105.L220201}. To be able to capture and digest this variety of phenomena, this work also introduces a powerful band‑structure analysis tool based on a commensurate approximation to the quasiperiodic potential, enabling us to precisely locate the boundaries separating localized, extended, and mixed phases using only unit cells of modest sizes.

	\textit{Model and methods.---}We consider a quasiperiodic ladder with a lateral shift between the upper and lower legs, of which the Hamiltonian can be written as, 
	\begin{eqnarray}
		H &= &t\sum_n \bigl(A^\dag_{n+1}A_n + B^\dag_{n+1}B_n\bigr) + \kappa\sum_n A^\dag_n B_n + \text{H.c.} \nonumber  \\
		&& + \sum_n \bigl(V_{n+q}A^\dag_n A_n + V_n B^\dag_n B_n\bigr).  
		\label{eq1}
	\end{eqnarray}
	Here, \(A_n\) and \(B_n\) denote the annihilation operators at site \(n\) of the upper and lower legs, respectively. The intraleg nearest-neighbor coupling is \(t\), and the interleg coupling is \(\kappa\). A quasiperiodic potential \(V_n = v \cos(2\pi\alpha n)\) is applied to the lower leg and with the upper leg shifted by $q$ sites,  the upper leg sees a shifted quasiperiodic potential $V_{n+q}$. Here, the parameter \(\alpha\) is chosen as the inverse golden ratio \((\sqrt{5}-1)/2\). In numerical calculations, the irrational $\alpha$ can be approximated by ratios of consecutive Fibonacci numbers \(F_n\), defined recursively by \(F_{n+1} = F_{n-1} + F_n\) with \(F_1 = F_2 = 1\). Specifically, \(\alpha = \lim_{n \to \infty} F_{n-1}/F_n\).  To facilitate our treatments below based on the periodic boundary condition, we shall use the rational approximation \(\alpha \approx F_{n-1}/F_n\) and then  take the system size \(L\) to be a Fibonacci number \(F_n\).  For example, Fig.~\ref{model}(a) illustrates the lateral-shifted quasiperiodic ladder obtained from this approximation with \(L = F_6=8\). 
		Under such a rational approximation,  $V_{n+q} = v \cos\big[2\pi (n+q) F_{n-1} / F_n\big]$. Since $F_{n-1}$ and $F_n$ are coprime, the expression $2\pi q F_{n-1} / F_n$ for $q=1,2,\dots$ can always be rewritten as $2\pi p' + 2\pi q' / F_n$, where $p'$ and $q'$ are integers and $q'$ runs over all values from $1$ to $F_n$ when $q$ is scanned~\cite{Suppl}.  As such,  the potential on the upper leg can be equivalently written as $V_{n+q} = v\cos(2\pi \alpha n + \phi)$, with $\phi = 2\pi q' / F_n$. The phase $\phi$ defined here becomes continuous in the limit of $n \to \infty$.

To proceed, we apply the duality transformation~\cite{PhysRevB.103.174205,SciPostPhys.13.3.046} \(A_n = \sum_m a_m e^{i2\pi\alpha m n}\) and \(B_n = \sum_m b_m e^{i2\pi\alpha m n}\) to Eq.~(\ref{eq1}). We then obtain the following Hamiltonian (in representation of the reciprocal momentum space)
\begin{eqnarray}
	H &=& \frac{v}{2}\sum_m \bigl(e^{i\phi}a^\dag_{m+1}a_m + b^\dag_{m+1}b_m\bigr) 
	+ \kappa\sum_m a^\dag_m b_m  \nonumber \\
	&& + \text{H.c.}+ \sum_m  2t \cos(2\pi\alpha m)  \bigl(a^\dag_m a_m + b^\dag_m b_m\bigr).
	\label{eq2}
\end{eqnarray}
Still viewing this duality-transformed Hamiltonian as a ladder system, then the intraleg coupling strength becomes \(v/2\), and the hopping on the upper leg acquires a Peierls phase \(\phi\), reflecting a gauge coupling (similar results can be obtained in off-diagonal modulated quasiperiodic ladders~\cite{Suppl}). The same onsite potential now applies to the two legs, but it is still quasiperiodic and is given by $2t \cos(2\pi\alpha m)$. Fig.~\ref{model}(b) illustrates an example of this duality-transformed Hamiltonian under the rational approximation with system size \(L = F_6=8\). Notably, each plaquette now encloses a magnetic flux \(\phi\), indicating that the lateral shift parameter \(q\) in Eq.~(\ref{eq1}) is equivalent to an effective magnetic field when viewed in reciprocal space, as evidenced by Eq.~(\ref{eq2}). Indeed, such a momentum-space magnetic field itself is of wide interest and has been realized in a dynamical setting using various platforms including cold atoms~\cite{PhysRevA.92.043606,sciadv.1602685,liang2024chiral}, superradiance lattices~\cite{PhysRevLett.122.023601} and the quantum kicked rotor~\cite{hainaut2018controlling}. Having a controllable effective magnetic flux in the reciprocal momentum space also strongly hints that the lateral shift is now a critical parameter to control the quasiperiodic ladder system. More importantly, none of the model Hamiltonians in Eqs.~(\ref{eq1}) and (\ref{eq2}) have been studied before, offering a fruitful angle to investigate localization-delocalization transitions in 1D. For subsequent analysis, we adopt \(v/2 = 1\) as the unit of energy and focus on the Hamiltonian given in Eq.~(\ref{eq2}), so that the localization-delocalization physics discussed below refers to that in the momentum space.  The model described by Eq.~(\ref{eq1}) is only the duality counterpart of Eq.~(\ref{eq2}), and hence exhibits completely opposite localization properties compared to those predicted from Eq.~(\ref{eq2}).

To characterize the localization properties of the model in Eq.~(\ref{eq2}), we introduce the inverse participation ratio (IPR) and the normalized participation ratio (NPR). For the $i$th eigenstate, $|\Psi_i\rangle = \sum_{m=1}^{L} (\psi^a_{i,m} a^{\dag}_{m} +\psi^b_{i,m} b^{\dag}_{m})|\mathrm{vac}\rangle$, its IPR is defined as $\mathrm{IPR}_i = \sum_{m=1}^{L} ( |\psi^a_{i,m}|^4 + |\psi^b_{i,m}|^4 )$ and the corresponding NPR is $\mathrm{NPR}_i = {1}/{(2L \cdot \mathrm{IPR}_i)}$. From these, we define the fractal dimension (FD) for the state as $\mathrm{FD}_i = -\lim_{L \to \infty} [{\ln \mathrm{IPR}_i}/{\ln (2L)}]$, which approaches 0 for localized states and 1 for extended states. To further discriminate mixed phases from purely localized or extended phases, we use the quantity $\eta = \log_{10}[ \langle \mathrm{IPR} \rangle \, \langle \mathrm{NPR} \rangle ]$, where $\langle \mathrm{IPR} \rangle$ and $\langle \mathrm{NPR} \rangle$ denote averages over all eigenstates for a fixed set of system parameters~\cite{PhysRevLett.126.106803,PhysRevB.101.064203}. In purely localized phases, $\langle \mathrm{IPR} \rangle$ approaches $1$ and $\langle \mathrm{NPR} \rangle $ approaches $ 1/(2L)$, whereas in purely extended phases $\langle \mathrm{IPR} \rangle $ approaches $1/(2L)$ and $\langle \mathrm{NPR} \rangle $ approaches $ 1$. In both phases, $\eta $ approaches $-\log_{10}(2L)$, which is lower than its value in mixed phases where $\langle \mathrm{IPR} \rangle$ and $\langle \mathrm{NPR} \rangle$ take intermediate values. Thus, $\eta$ serves as a diagnostic to clearly distinguish extended or localized phases from mixed ones~\cite{PhysRevLett.126.106803,PhysRevB.101.064203}.

\begin{figure}
	\includegraphics[width=\linewidth]{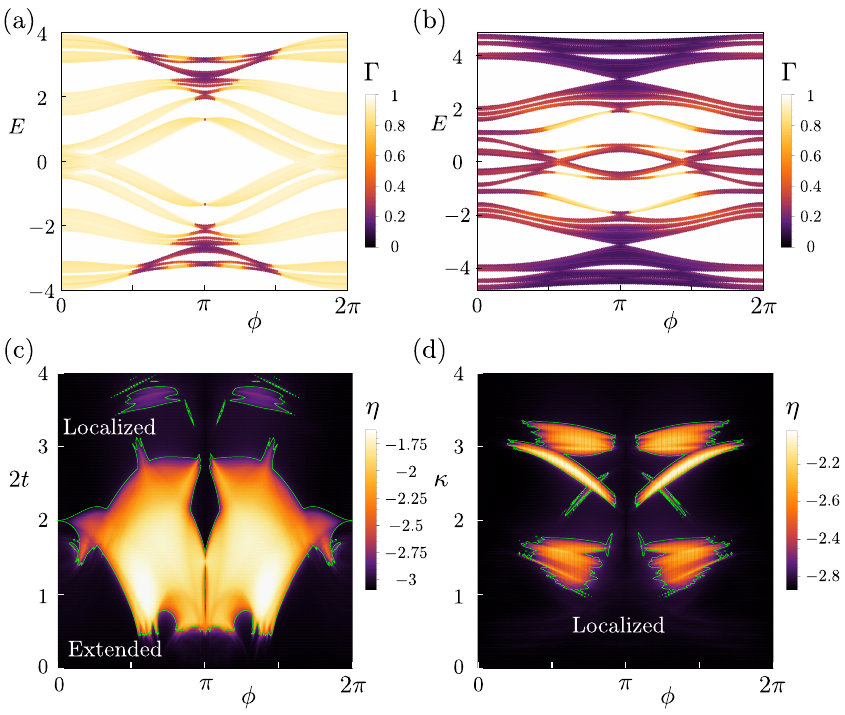}
	\caption{Flux-dependent localization. (a) Energy spectrum vs the effective magnetic flux $\phi$, plotted with the fractal dimension $\Gamma$ of the eigenstates, for system parameters $\kappa=1.8$ and $2t=1$. (b) Same as in (a) but for $2t=2.5$. (c) Phase diagram vs the quasiperiodic potential strength $2t$ and  $\phi$, with $\kappa=1.8$, color-coded by the values of $\eta$. (d) Same as in (c), but for $\kappa$ and  $\phi$, with $2t=3.4$. The system size is taken as $L = F_{15} = 610$. Green lines in panels (c) and (d) represent the theoretical analysis, in excellent agreement with the numerically found phase boundaries.}
	\label{MobilityEdge}
\end{figure}

\textit{Localized, extended and mixed phases in 1D.---} Figure~\ref{MobilityEdge}(a) presents the calculated fractal dimension $\Gamma$ of each eigenstate vs its respective eigenenergy, as a function of the effective magnetic flux $\phi$ for parameters $\kappa=1.8$ and $2t=1$, with $L=F_{15}=610$. For such a system size, tuning the lateral shift $q$ will essentially lead to a continuous tuning in $\phi$. At $\phi=0$, all states are extended ($\Gamma\simeq 1$), reflecting a relatively weak quasiperiodic potential as our starting point. As $\phi$ increases towards $\pi/2$, the eigenstates begin to localize from the upper and lower band edges. This is a magnetic-flux-enhanced localization. 
Figure~\ref{MobilityEdge}(b) illustrates the opposite scenario with $\kappa=1.8$ and $2t=2.5$. There, at $\phi=0$, the stronger quasiperiodic potential strength (being much larger than 2) has caused all the eigenstates to be localized, with the fractal dimension $\Gamma\simeq 0$.  Upon increasing $\phi$ to around $\pi/2$, extended states emerge from the band center near $E=0$.  This indicates that the effective magnetic flux can be used to suppress localization. 

Figure~\ref{MobilityEdge}(c) presents the quantity $\eta$ defined above versus $2t$ and $\phi$ for $\kappa=1.8$. The black regions, corresponding to the  smallest values of $\eta$, mark purely localized or purely extended phases (as labeled). The bright regions, where $\eta$ takes larger values, correspond to mixed phases where localized and extended states coexist.
Figure~\ref{MobilityEdge}(d) depicts $\eta$ vs $\kappa$ and $\phi$ for $2t=3.4$. Again, the plotted black regions mark purely localized phases, whereas the bright regions signal the mixed phases. From Figs.~\ref{MobilityEdge}(c) and~\ref{MobilityEdge}(d), we observe intriguing reentrant localization transitions over a broad range of $\phi$ when $t$ or $\kappa$ is tuned.

As a starting point to understand the results in Fig.~\ref{MobilityEdge}, we note that for $\phi=0$ or $\phi=\pi$, the duality-transformed quasiperiodic ladder can be viewed, after a simple rotation of basis states,  as two decoupled AA models when $\phi=0$ or the same but with additional staggered onsite potential when $\phi=\pi$ (see Supplementary Materials~\cite{Suppl} for technical details), with the latter case known to support reentrant localization transitions.  With this insight, for $\phi=0$, we expect that the localization properties of the system are independent of $\kappa$ and only depend on the quasiperiodic potential strength $2t$~\cite{aubry1980analyticity}. For $|2t|<2$, all the eigenstates are extended [$\phi=0$ in Fig.~\ref{MobilityEdge}(a)]. For  $|2t|>2$, all the eigenstates are localized [$\phi=0$ of Fig.~\ref{MobilityEdge}(b)], with $|2t|=2$ being the phase transition point [$\phi=0$ in Fig.~\ref{MobilityEdge}(c)].  For the case of $\phi=\pi$, the two decoupled equations correspond each to an AA model with an additional staggered onsite potential~\cite{PhysRevLett.126.106803,PhysRevB.105.L220201}. As also shown in the Supplementary Materials~\cite{Suppl}, the quasiperiodic ladders for $\phi=\pi$ exhibit the same phase diagram as that reported in Refs.~\cite{PhysRevLett.126.106803,PhysRevB.105.L220201}. Notably, that the lateral shift parameter $\phi$ can include but go far beyond these two special cases motivates a systematic analysis to better understand the localization-delocalization transitions in our 1D ladder system.

Note in passing that the green curves in both Figs.~\ref{MobilityEdge}(c) and \ref{MobilityEdge}(d) feature the theoretically obtained phase boundaries between the mixed phases and the purely localized or extended phases, in agreement with our computational findings using the $\eta$ indicator defined above. We elaborate on our theoretical analysis below.

\textit{Band structure analysis based on a commensurate approximation.—}Consider now a different gauge transformation \(a_m \rightarrow \tilde a_m e^{i m \phi/2}\) and \(b_m \rightarrow \tilde b_m e^{i m \phi/2}\) applied to Eq.~(\ref{eq2}), we then obtain the following 1D Hamiltonian  
\begin{eqnarray}  
	\tilde H &= &\sum_m \Bigl(e^{i\phi/2} \tilde a^\dag_{m+1} \tilde a_m + e^{-i\phi/2} \tilde b^\dag_{m+1} \tilde b_m\Bigr) +  \kappa\sum_m \tilde a^\dag_m \tilde b_m \nonumber \\  
	&& + \mathrm{H.c.}+ \sum_m  2t\cos(2\pi\alpha m)  \bigl(\tilde a^\dag_m \tilde a_m + \tilde b^\dag_m \tilde b_m\bigr)  .   
	\label{eq3}
\end{eqnarray}  
This Hamiltonian should be compared with the following 2D Hamiltonian for a bilayer lattice in the $xy$ plane in a magnetic field, i.e.,
\begin{widetext}
	\begin{equation}
		\tilde H_{xy} = \sum_{m,n} \Bigl(e^{i\phi/2} \tilde a^\dag_{m+1,n} \tilde a_{m,n}  
		+ e^{-i\phi/2} \tilde b^\dag_{m+1,n} \tilde b_{m,n}\Bigr) +\kappa\sum_{m,n} \tilde a^\dag_{m,n} \tilde b_{m,n}   
		+ \sum_{m,n} t e^{i 2\pi\alpha m}  
		\bigl(\tilde a^\dag_{m,n+1} \tilde a_{m,n} + \tilde b^\dag_{m,n+1} \tilde b_{m,n}\bigr) +  \mathrm{H.c.},
		\label{eq6}
	\end{equation}
\end{widetext}
subject to a uniform magnetic flux \(2\pi\alpha\) along the \(z\)-direction and \(\phi\) along the \(y\)-direction, with
the hopping amplitudes being unity along the \(x\)-direction and $t$ along the \(y\)-direction, and the interlayer coupling \(\kappa\).
Indeed, performing a partial Fourier transform on \(\tilde H_{xy}\) in the \(y\)-direction would reduce it to the 1D form \(\tilde H\) (with $k_y=0$), indicating a direct analogy between our quasiperiodic ladder and the standard Hofstadter model~\cite{PhysRevB.14.2239}.  

We now advocate to make use of $\tilde H_{xy}$ as the parent 2D system to systematically investigate the localization-delocalization transitions in our ladder system, by first applying a commensurate approximation to \(\alpha\), taking \(\alpha \approx F_{n-1}/F_n\). The 2D Hamiltonian in Eq.~(\ref{eq6}) then becomes periodic with lattice vectors \(\vec l = (l_{\mathrm{CA}}, l_y) = (F_n, 1)\) and block-diagonalizes as \(\tilde H_{xy} = \sum_{k_x,k_y} \tilde H(k_x,k_y)\), yielding
energy bands \(E_j(k_x,k_y)\) 
with $j$ the band index. 
There is also a spectral reflection symmetry for  even  \(l_{\mathrm{CA}}\), as proved in the Supplementary Materials~\cite{Suppl}.  

\begin{figure}
	\includegraphics[width=\linewidth]{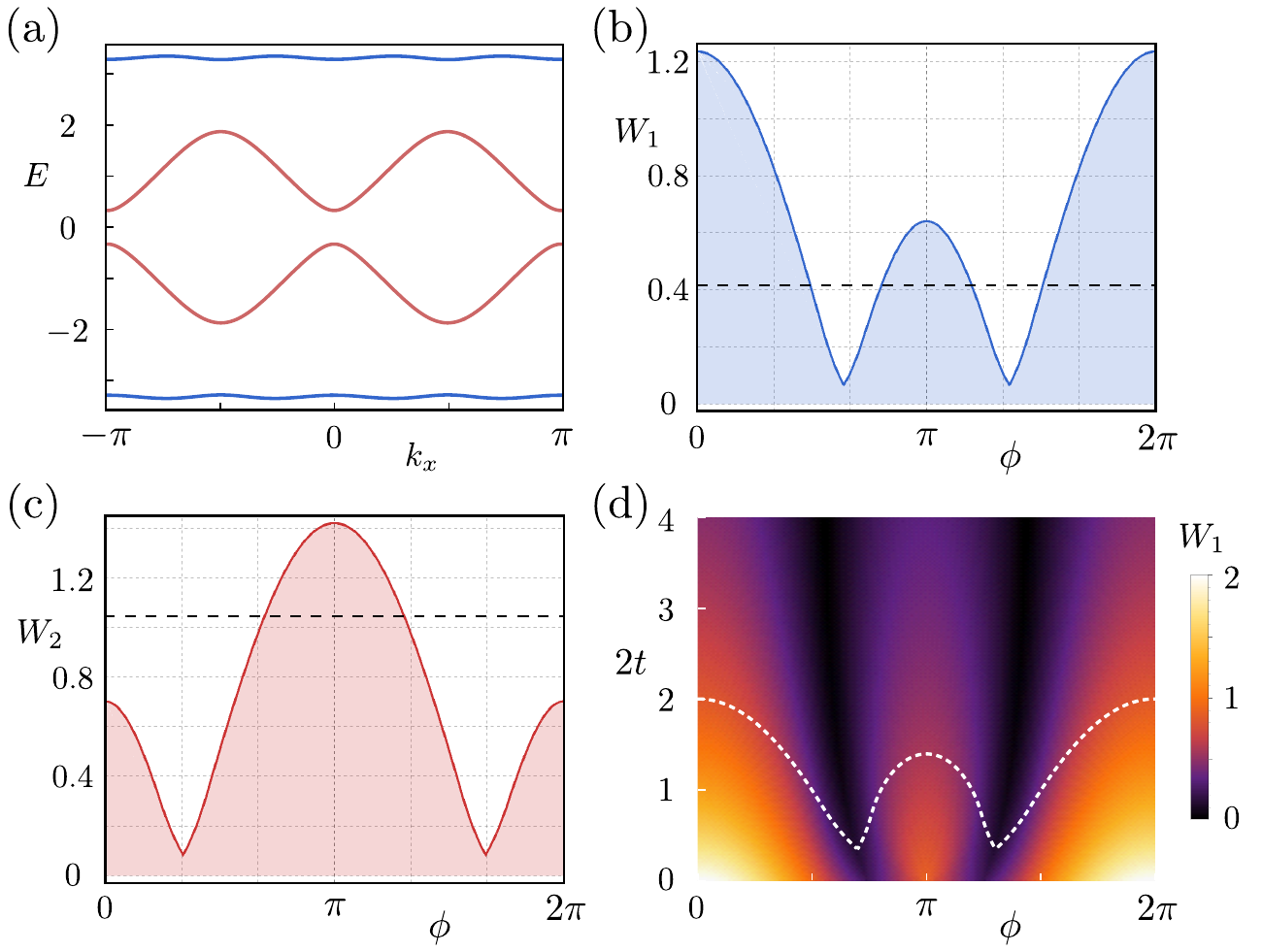}
	\caption{Bandwidth under a commensurate approximation,  with lattice constant \GJ{\(l_{\mathrm{CA}} = F_3 = 2\)}. (a) Band structure for $\kappa=1.8$, $2t=1$ and $\phi=2$.
		(b) Bandwidth of the first band \(W_1\) vs \(\phi\) for \(\kappa = 1.8\) and \(2t = 1\), with other parameters the same as in Fig.~\ref{MobilityEdge}(a).
		(c) Bandwidth of the second band \(W_2\) vs \(\phi\) for \(\kappa = 1.8\) and \(2t = 2.5\), with other system parameters the same as in Fig.~\ref{MobilityEdge}(b).
		Dashed lines in panels (b) and (c) indicate where the bandwidth ($W_1$ and $W_2$, respectively) equals the renormalized strength of quasiperiodic potential $2\tilde{t} = 2(\sqrt{2} - 1)t$.
		(d) Bandwidth \(W_1\) vs \(\phi\) and \(2t\) for \(\kappa = 1.8\). White dashed line marks the condition \(W_1 = 2\tilde{t}\).}
	\label{bandwidth}
\end{figure}


\begin{figure*}
	\includegraphics[width=\linewidth]{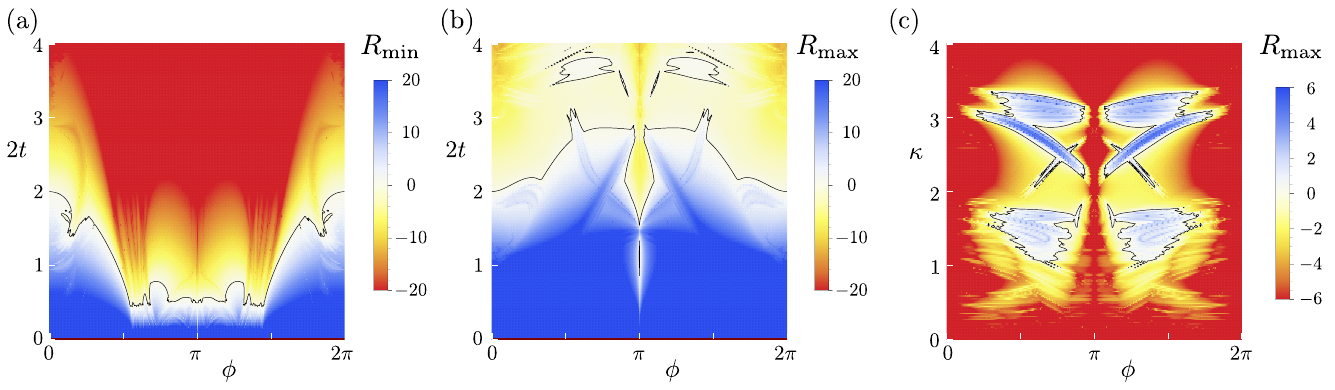}
	\caption{Anisotropic bandwidth ratio obtained under a commensurate approximation with  \GJ{\(l_{\mathrm{CA}} = F_9 = 34\).}
		(a) Logarithmic map of the minimum anisotropic bandwidth ratio $R_{\min}$  vs \(2t\) and \(\phi\) for \(\kappa = 1.8\) as in Fig.~\ref{MobilityEdge}(c).
		(b) Same as in (a), but for the maximal anisotropic bandwidth ratio $R_{\max}$ vs  \(2t\) and \(\phi\).
		(c) Same as in (b), but for  $R_{\max}$ vs  \(\kappa\) and \(\phi\) for \(2t = 3.4\) as in Fig.~\ref{MobilityEdge}(d).
		Black curves indicate where \(R_{\min}=1\) or \(R_{\max}=1\), plotted as green curves in Figs.~\ref{MobilityEdge}(c) and~\ref{MobilityEdge}(d).}
	\label{txty}
\end{figure*}

Under the commensurate approximation with lattice constant \(l_{\text{CA}}=F_3=2\), Fig.~\ref{bandwidth}(a) presents the corresponding band structure for \(\kappa=1.8\), \(2t=1\), and \(\phi=2\), with $k_y$ restricted to $k_y=0$. Four bands appear, with the upper and lower bands being quasi-flat. Such flat bands originate from the interplay between the magnetic flux and onsite potential modulation--a mechanism also observed in chiral ladders with modulated interleg couplings~\cite{PhysRevB.102.155429}. Their formation shares characteristics similar to those of magic-angle twisted bilayer graphene~\cite{PhysRevLett.99.256802,Trambly2010,pnas.1108174108}. Importantly, this flat-band formation underlies the magnetic-flux-enhanced localization seen in Fig.~\ref{MobilityEdge}(a). In Fig.~\ref{bandwidth}(b), we plot the bandwidth \(W_1\) of the first band as a function of \(\phi\) for \(\kappa=1.8\) and \(2t=1\), the same system parameters used in Fig.~\ref{MobilityEdge}(a). As \(\phi\) increases from \(0\) to \(\pi\), the bandwidth first decreases and then increases. The dashed line marks the position of the renormalized quasiperiodic potential strength \(2\tilde{t}\) elaborated in the Supplementary Materials~\cite{Suppl}. The introduction of renormalized quasiperiodic potential strengths is appealing because, when \(2\tilde{t} < W_1\), eigenstates in the band remain extended; and when \(2\tilde{t} > W_1\), they become localized. Thus, assisted by the introduction of $2\tilde{t}$, the bandwidth \(W_1\) in Fig.~\ref{bandwidth}(b) vs $2\tilde{t}$ provides an excellent description of the localization features presented in Fig.~\ref{MobilityEdge}(a), in a way much similar to the AA model.

Notably, we can now also explain the observed magnetic-flux-suppressed localization in Fig.~\ref{MobilityEdge}(b) in terms of the broadening of the middle bands near \(E=0\). Figure~\ref{bandwidth}(c) displays the bandwidth \(W_2\) of the second band vs \(\phi\) for \(\kappa = 1.8\) and \(2t = 2.5\), matching the parameters used in Fig.~\ref{MobilityEdge}(b).  For \(\phi=0\), we find \(2\tilde{t} > W_2\), and the corresponding states are localized. As \(\phi\) increases toward \(\pi\), \(2\tilde{t}\) becomes smaller than \(W_2\), and the states become extended. Hence, the behavior of \(W_2\) in Fig.~\ref{bandwidth}(c) vs $2\tilde{t}$ does explain the localization and delocalization features observed in Fig.~\ref{MobilityEdge}(b).

In Fig.~\ref{bandwidth}(d), we examine \(W_1\) of the first band as a function of both the effective magnetic flux \(\phi\) and the quasiperiodic potential strength \(2t\) for \(\kappa = 1.8\). The white dashed line marks where \(W_1 = 2\tilde{t}\). Below this line, where \(2\tilde{t} < W_1<W_2\), all states remain extended. Above it, where \(2\tilde{t} > W_1\) but still with \(2\tilde{t} < W_2\), states from the band edges begin to localize.  As such, the dashed line in Fig.~\ref{bandwidth}(d) offers an approximate boundary between extended and mixed phases, consistent with the phase diagram in Fig.~\ref{MobilityEdge}(c).
In the Supplementary Materials~\cite{Suppl}, we further confirm the physical mechanism for other large values of $\kappa$: the magnetic-flux-enhanced localization always stems from bandwidth narrowing near the band edges, whereas the suppression of localization arises from bandwidth broadening around the band center. 

The band narrowing or widening of the 2D parent Hamiltonian $\tilde H_{xy}$ (using a very small unit cell) has already successfully captured the localization-delocalization features in Fig.~\ref{MobilityEdge}(c).  \GJ{This fact motivates us to further leverage the relation between the 1D Hamiltonian $\tilde{H}$  in Eq.~(\ref{eq3}) and the elevated 2D Hamiltonian  $\tilde{H}_{xy}$ in Eq.~(\ref{eq6}).   For the 1D system $\tilde{H}$, the ratio of the quasiperiodic potential strength (be it bare or renormalized) with the band width along $k_x$ is critical to predict the localization-delocalization along the $x$ direction. This suggests that
	for $\tilde{H}_{xy}$, if treated under a commensurate approximation to the \WW{quasiperiodic potential},  we should quantitatively examine the bandwidth ratio along $k_x$ with that along $k_y$ (naturally replacing the quasiperiodic potential strength), thus determining the phase boundaries along the $x$ dimension,  separating purely localized or delocalized phases from mixed phases}.  To that end, we now introduce the anisotropic bandwidth ratio, 
by making use of the full band structure \(E_j(k_x,k_y)\).  Specifically,  we let the bandwidth along the \(k_x\)- and \(k_y\)-directions be  $W_{x, j}=\max\limits_{k_y}\Bigl[ \max\limits_{k_x}E_j(k_x,k_y)-\min\limits_{k_x}E_j(k_x,k_y)\Bigr]$ and
$W_{y, j}=\max\limits_{k_x}\Bigl[ \max\limits_{k_y}E_j(k_x,k_y)-\min\limits_{k_y}E_j(k_x,k_y)\Bigr]$. 
The anisotropic bandwidth ratio for the $j$th band is then defined to be \(R_j=W_{x, j}/W_{y, j}\).  As observed below, for rather modest-sized unit cells, 
\(R_j>1\) can already precisely predict that the states on the $j$th band are extended along \(x\), whereas \(R_j<1\) predicts localization. This is computationally appealing, because there is no longer a need to handle very large lattices.   Compared with other localization indicators such as the fractal dimension $\Gamma$ or the $\eta$ parameter, $R_j$ defined here provides a much sharper signature of the localization-delocalization phase transition. The critical situation \(R_j=1\), confirmed by all our numerical studies, serves as a system-size-independent threshold, in contrast to the other measures $\Gamma$ and $\eta$ used previously, whose values at the transition point vary with both system sizes and model details.

Taking \(l_{\mathrm{CA}} = F_9 = 34\) as an example, in Fig.~\ref{txty}(a) we present the logarithmic map of the minimum anisotropic bandwidth ratio over all bands, \({R}_{\min} = \min_j R_j\), as a function of \(2t\) and \(\phi\) for \(\kappa = 1.8\) [as in Fig.~\ref{MobilityEdge}(c)]. Because 
\(R_{\min}\) characterizes the band most prone to localization, the condition \(R_{\min}=1\) marks the phase boundary between purely extended and mixed phases. Figure~\ref{txty}(b) shows the logarithmic map of the maximum ratio, \(R_{\max} = \max_j R_j\), for the same \(\kappa\). Here \(R_{\max}=1\) identifies the boundary between purely localized and mixed phases. The black curves in Figs.~\ref{txty}(a) and~\ref{txty}(b) indicate where \(R_{\min}=1\) and \(R_{\max}=1\), respectively, plotted as green curves in Fig.~\ref{MobilityEdge}(c).  Our introduced anisotropic bandwidth ratio is  seen to accurately capture the phase boundaries identified in Fig.~\ref{MobilityEdge}(c). Finally, Fig.~\ref{txty}(c) displays the logarithmic map of \(R_{\max}\) vs \(\kappa\) and \(\phi\) for \(2t = 3.4\) as in Fig.~\ref{MobilityEdge}(d). The black curve marks \(R_{\max}=1\), delineating the transition from purely localized to mixed phases. When superimposed as green curves in Fig.~\ref{MobilityEdge}(d), this line again matches perfectly the phase boundary obtained from other measures such as $\eta$.

\textit{Conclusion.---} The integration of synthetic magnetic flux with quasiperiodicity was previously known to induce rich localization phenomena~\cite{PhysRevX.8.031045},  yet the complexity of realizing the required magnetic flux limits their relevance to a broader range of systems.  This work revives this topic by showing that a simple lateral shift between the upper and lower legs of a quasiperiodic 1D ladder can synthesize a magnetic flux in the reciprocal lattice and hence extensively control  the localization-delocalization transitions. 
The underlying physics induced by such lateral shifts should be readily realizable across a range of experimental platforms~\cite{Wiersma,PhysRevLett.112.146404,Wang2020,PhysRevLett.125.133603,XIAO20212175,PhysRevLett.129.103401,PhysRevX.11.011016,PhysRevLett.129.173601,PhysRevLett.132.066602,PhysRevLett.134.053601,Hou2026}. This would enable the observation of magnetic-flux-enhanced localization, magnetic-flux-suppressed localization, and magnetic-flux-induced reentrant localization transitions under conditions much broader than those previously accessible.
Qualitatively, the spectral bandwidth narrowing or broadening is the underlying mechanism activated by the interplay of the synthetic magnetic flux and the quasiperiodic potential. We have also revealed that the anisotropic bandwidth ratio, obtained from band structure analyses using modest-sized unit cells under some commensurate approximations, suffices to capture the phase boundaries separating regions of localized,  extended, and mixed phases. 

In future work, our insight that a simple lateral shift can serve as a synthetic magnetic flux in the reciprocal momentum space may be used to manipulate other lattice systems, such as purely periodic or entirely disordered systems in 1D and 2D. Furthermore, our unique methodology based on the anisotropic bandwidth ratio can be fruitfully applied to other systems, including non‑Hermitian topological localization transitions~\cite{PhysRevLett.122.237601,PhysRevB.100.054301}.


\section{Acknowledgments}
W.Z. is supported by the National Natural Science Foundation of China (Grants No.~12404499). L.Z. is supported by the National Natural Science Foundation of China (Grants No.~12275260 and No.~11905211) and the Young Talents Project of Ocean University of China. J.G. acknowledges support by the National Research Foundation, Singapore, through the National Quantum Office, hosted in A*STAR, under its Centre for Quantum Technologies Funding Initiative (S24Q2d0009).


\begin{thebibliography}{80}%
	\makeatletter
	\providecommand \@ifxundefined [1]{%
		\@ifx{#1\undefined}
	}%
	\providecommand \@ifnum [1]{%
		\ifnum #1\expandafter \@firstoftwo
		\else \expandafter \@secondoftwo
		\fi
	}%
	\providecommand \@ifx [1]{%
		\ifx #1\expandafter \@firstoftwo
		\else \expandafter \@secondoftwo
		\fi
	}%
	\providecommand \natexlab [1]{#1}%
	\providecommand \enquote  [1]{``#1''}%
	\providecommand \bibnamefont  [1]{#1}%
	\providecommand \bibfnamefont [1]{#1}%
	\providecommand \citenamefont [1]{#1}%
	\providecommand \href@noop [0]{\@secondoftwo}%
	\providecommand \href [0]{\begingroup \@sanitize@url \@href}%
	\providecommand \@href[1]{\@@startlink{#1}\@@href}%
	\providecommand \@@href[1]{\endgroup#1\@@endlink}%
	\providecommand \@sanitize@url [0]{\catcode `\\12\catcode `\$12\catcode
		`\&12\catcode `\#12\catcode `\^12\catcode `\_12\catcode `\%12\relax}%
	\providecommand \@@startlink[1]{}%
	\providecommand \@@endlink[0]{}%
	\providecommand \url  [0]{\begingroup\@sanitize@url \@url }%
	\providecommand \@url [1]{\endgroup\@href {#1}{\urlprefix }}%
	\providecommand \urlprefix  [0]{URL }%
	\providecommand \Eprint [0]{\href }%
	\providecommand \doibase [0]{https://doi.org/}%
	\providecommand \selectlanguage [0]{\@gobble}%
	\providecommand \bibinfo  [0]{\@secondoftwo}%
	\providecommand \bibfield  [0]{\@secondoftwo}%
	\providecommand \translation [1]{[#1]}%
	\providecommand \BibitemOpen [0]{}%
	\providecommand \bibitemStop [0]{}%
	\providecommand \bibitemNoStop [0]{.\EOS\space}%
	\providecommand \EOS [0]{\spacefactor3000\relax}%
	\providecommand \BibitemShut  [1]{\csname bibitem#1\endcsname}%
	\let\auto@bib@innerbib\@empty
	\bibitem [{\citenamefont {Anderson}(1958)}]{PhysRev.109.1492}%
	\BibitemOpen
	\bibfield  {author} {\bibinfo {author} {\bibfnamefont {P.~W.}\ \bibnamefont
			{Anderson}},\ }\bibfield  {title} {\bibinfo {title} {Absence of diffusion in
			certain random lattices},\ }\href {https://doi.org/10.1103/PhysRev.109.1492}
	{\bibfield  {journal} {\bibinfo  {journal} {Phys. Rev.}\ }\textbf {\bibinfo
			{volume} {109}},\ \bibinfo {pages} {1492} (\bibinfo {year}
		{1958})}\BibitemShut {NoStop}%
	\bibitem [{\citenamefont {Evers}\ and\ \citenamefont
		{Mirlin}(2008)}]{RevModPhys.80.1355}%
	\BibitemOpen
	\bibfield  {author} {\bibinfo {author} {\bibfnamefont {F.}~\bibnamefont
			{Evers}}\ and\ \bibinfo {author} {\bibfnamefont {A.~D.}\ \bibnamefont
			{Mirlin}},\ }\bibfield  {title} {\bibinfo {title} {Anderson transitions},\
	}\href {https://doi.org/10.1103/RevModPhys.80.1355} {\bibfield  {journal}
		{\bibinfo  {journal} {Rev. Mod. Phys.}\ }\textbf {\bibinfo {volume} {80}},\
		\bibinfo {pages} {1355} (\bibinfo {year} {2008})}\BibitemShut {NoStop}%
	\bibitem [{\citenamefont {Abrahams}\ \emph {et~al.}(1979)\citenamefont
		{Abrahams}, \citenamefont {Anderson}, \citenamefont {Licciardello},\ and\
		\citenamefont {Ramakrishnan}}]{PhysRevLett.42.673}%
	\BibitemOpen
	\bibfield  {author} {\bibinfo {author} {\bibfnamefont {E.}~\bibnamefont
			{Abrahams}}, \bibinfo {author} {\bibfnamefont {P.~W.}\ \bibnamefont
			{Anderson}}, \bibinfo {author} {\bibfnamefont {D.~C.}\ \bibnamefont
			{Licciardello}},\ and\ \bibinfo {author} {\bibfnamefont {T.~V.}\ \bibnamefont
			{Ramakrishnan}},\ }\bibfield  {title} {\bibinfo {title} {Scaling theory of
			localization: Absence of quantum diffusion in two dimensions},\ }\href
	{https://doi.org/10.1103/PhysRevLett.42.673} {\bibfield  {journal} {\bibinfo
			{journal} {Phys. Rev. Lett.}\ }\textbf {\bibinfo {volume} {42}},\ \bibinfo
		{pages} {673} (\bibinfo {year} {1979})}\BibitemShut {NoStop}%
	\bibitem [{\citenamefont {Das~Sarma}\ \emph {et~al.}(1988)\citenamefont
		{Das~Sarma}, \citenamefont {He},\ and\ \citenamefont
		{Xie}}]{PhysRevLett.61.2144}%
	\BibitemOpen
	\bibfield  {author} {\bibinfo {author} {\bibfnamefont {S.}~\bibnamefont
			{Das~Sarma}}, \bibinfo {author} {\bibfnamefont {S.}~\bibnamefont {He}},\ and\
		\bibinfo {author} {\bibfnamefont {X.~C.}\ \bibnamefont {Xie}},\ }\bibfield
	{title} {\bibinfo {title} {Mobility edge in a model one-dimensional
			potential},\ }\href {https://doi.org/10.1103/PhysRevLett.61.2144} {\bibfield
		{journal} {\bibinfo  {journal} {Phys. Rev. Lett.}\ }\textbf {\bibinfo
			{volume} {61}},\ \bibinfo {pages} {2144} (\bibinfo {year}
		{1988})}\BibitemShut {NoStop}%
	\bibitem [{\citenamefont {Boers}\ \emph {et~al.}(2007)\citenamefont {Boers},
		\citenamefont {Goedeke}, \citenamefont {Hinrichs},\ and\ \citenamefont
		{Holthaus}}]{PhysRevA.75.063404}%
	\BibitemOpen
	\bibfield  {author} {\bibinfo {author} {\bibfnamefont {D.~J.}\ \bibnamefont
			{Boers}}, \bibinfo {author} {\bibfnamefont {B.}~\bibnamefont {Goedeke}},
		\bibinfo {author} {\bibfnamefont {D.}~\bibnamefont {Hinrichs}},\ and\
		\bibinfo {author} {\bibfnamefont {M.}~\bibnamefont {Holthaus}},\ }\bibfield
	{title} {\bibinfo {title} {Mobility edges in bichromatic optical lattices},\
	}\href {https://doi.org/10.1103/PhysRevA.75.063404} {\bibfield  {journal}
		{\bibinfo  {journal} {Phys. Rev. A}\ }\textbf {\bibinfo {volume} {75}},\
		\bibinfo {pages} {063404} (\bibinfo {year} {2007})}\BibitemShut {NoStop}%
	\bibitem [{\citenamefont {Biddle}\ \emph {et~al.}(2009)\citenamefont {Biddle},
		\citenamefont {Wang}, \citenamefont {Priour},\ and\ \citenamefont
		{Das~Sarma}}]{PhysRevA.80.021603}%
	\BibitemOpen
	\bibfield  {author} {\bibinfo {author} {\bibfnamefont {J.}~\bibnamefont
			{Biddle}}, \bibinfo {author} {\bibfnamefont {B.}~\bibnamefont {Wang}},
		\bibinfo {author} {\bibfnamefont {D.~J.}\ \bibnamefont {Priour}},\ and\
		\bibinfo {author} {\bibfnamefont {S.}~\bibnamefont {Das~Sarma}},\ }\bibfield
	{title} {\bibinfo {title} {Localization in one-dimensional incommensurate
			lattices beyond the aubry-andr\'e model},\ }\href
	{https://doi.org/10.1103/PhysRevA.80.021603} {\bibfield  {journal} {\bibinfo
			{journal} {Phys. Rev. A}\ }\textbf {\bibinfo {volume} {80}},\ \bibinfo
		{pages} {021603} (\bibinfo {year} {2009})}\BibitemShut {NoStop}%
	\bibitem [{\citenamefont {Lellouch}\ and\ \citenamefont
		{Sanchez-Palencia}(2014)}]{PhysRevA.90.061602}%
	\BibitemOpen
	\bibfield  {author} {\bibinfo {author} {\bibfnamefont {S.}~\bibnamefont
			{Lellouch}}\ and\ \bibinfo {author} {\bibfnamefont {L.}~\bibnamefont
			{Sanchez-Palencia}},\ }\bibfield  {title} {\bibinfo {title} {Localization
			transition in weakly interacting bose superfluids in one-dimensional
			quasiperdiodic lattices},\ }\href
	{https://doi.org/10.1103/PhysRevA.90.061602} {\bibfield  {journal} {\bibinfo
			{journal} {Phys. Rev. A}\ }\textbf {\bibinfo {volume} {90}},\ \bibinfo
		{pages} {061602} (\bibinfo {year} {2014})}\BibitemShut {NoStop}%
	\bibitem [{\citenamefont {Liu}\ \emph {et~al.}(2015)\citenamefont {Liu},
		\citenamefont {Ghosh},\ and\ \citenamefont {Chong}}]{PhysRevB.91.014108}%
	\BibitemOpen
	\bibfield  {author} {\bibinfo {author} {\bibfnamefont {F.}~\bibnamefont
			{Liu}}, \bibinfo {author} {\bibfnamefont {S.}~\bibnamefont {Ghosh}},\ and\
		\bibinfo {author} {\bibfnamefont {Y.~D.}\ \bibnamefont {Chong}},\ }\bibfield
	{title} {\bibinfo {title} {Localization and adiabatic pumping in a
			generalized aubry-andr\'e-harper model},\ }\href
	{https://doi.org/10.1103/PhysRevB.91.014108} {\bibfield  {journal} {\bibinfo
			{journal} {Phys. Rev. B}\ }\textbf {\bibinfo {volume} {91}},\ \bibinfo
		{pages} {014108} (\bibinfo {year} {2015})}\BibitemShut {NoStop}%
	\bibitem [{\citenamefont {Bomantara}\ \emph {et~al.}(2016)\citenamefont
		{Bomantara}, \citenamefont {Raghava}, \citenamefont {Zhou},\ and\
		\citenamefont {Gong}}]{PhysRevE.93.022209}%
	\BibitemOpen
	\bibfield  {author} {\bibinfo {author} {\bibfnamefont {R.~W.}\ \bibnamefont
			{Bomantara}}, \bibinfo {author} {\bibfnamefont {G.~N.}\ \bibnamefont
			{Raghava}}, \bibinfo {author} {\bibfnamefont {L.}~\bibnamefont {Zhou}},\ and\
		\bibinfo {author} {\bibfnamefont {J.}~\bibnamefont {Gong}},\ }\bibfield
	{title} {\bibinfo {title} {Floquet topological semimetal phases of an
			extended kicked harper model},\ }\href
	{https://doi.org/10.1103/PhysRevE.93.022209} {\bibfield  {journal} {\bibinfo
			{journal} {Phys. Rev. E}\ }\textbf {\bibinfo {volume} {93}},\ \bibinfo
		{pages} {022209} (\bibinfo {year} {2016})}\BibitemShut {NoStop}%
	\bibitem [{\citenamefont {L\"uschen}\ \emph {et~al.}(2018)\citenamefont
		{L\"uschen}, \citenamefont {Scherg}, \citenamefont {Kohlert}, \citenamefont
		{Schreiber}, \citenamefont {Bordia}, \citenamefont {Li}, \citenamefont
		{Das~Sarma},\ and\ \citenamefont {Bloch}}]{PhysRevLett.120.160404}%
	\BibitemOpen
	\bibfield  {author} {\bibinfo {author} {\bibfnamefont {H.~P.}\ \bibnamefont
			{L\"uschen}}, \bibinfo {author} {\bibfnamefont {S.}~\bibnamefont {Scherg}},
		\bibinfo {author} {\bibfnamefont {T.}~\bibnamefont {Kohlert}}, \bibinfo
		{author} {\bibfnamefont {M.}~\bibnamefont {Schreiber}}, \bibinfo {author}
		{\bibfnamefont {P.}~\bibnamefont {Bordia}}, \bibinfo {author} {\bibfnamefont
			{X.}~\bibnamefont {Li}}, \bibinfo {author} {\bibfnamefont {S.}~\bibnamefont
			{Das~Sarma}},\ and\ \bibinfo {author} {\bibfnamefont {I.}~\bibnamefont
			{Bloch}},\ }\bibfield  {title} {\bibinfo {title} {Single-particle mobility
			edge in a one-dimensional quasiperiodic optical lattice},\ }\href
	{https://doi.org/10.1103/PhysRevLett.120.160404} {\bibfield  {journal}
		{\bibinfo  {journal} {Phys. Rev. Lett.}\ }\textbf {\bibinfo {volume} {120}},\
		\bibinfo {pages} {160404} (\bibinfo {year} {2018})}\BibitemShut {NoStop}%
	\bibitem [{\citenamefont {Longhi}(2019)}]{PhysRevLett.122.237601}%
	\BibitemOpen
	\bibfield  {author} {\bibinfo {author} {\bibfnamefont {S.}~\bibnamefont
			{Longhi}},\ }\bibfield  {title} {\bibinfo {title} {Topological phase
			transition in non-hermitian quasicrystals},\ }\href
	{https://doi.org/10.1103/PhysRevLett.122.237601} {\bibfield  {journal}
		{\bibinfo  {journal} {Phys. Rev. Lett.}\ }\textbf {\bibinfo {volume} {122}},\
		\bibinfo {pages} {237601} (\bibinfo {year} {2019})}\BibitemShut {NoStop}%
	\bibitem [{\citenamefont {Jiang}\ \emph {et~al.}(2019)\citenamefont {Jiang},
		\citenamefont {Lang}, \citenamefont {Yang}, \citenamefont {Zhu},\ and\
		\citenamefont {Chen}}]{PhysRevB.100.054301}%
	\BibitemOpen
	\bibfield  {author} {\bibinfo {author} {\bibfnamefont {H.}~\bibnamefont
			{Jiang}}, \bibinfo {author} {\bibfnamefont {L.-J.}\ \bibnamefont {Lang}},
		\bibinfo {author} {\bibfnamefont {C.}~\bibnamefont {Yang}}, \bibinfo {author}
		{\bibfnamefont {S.-L.}\ \bibnamefont {Zhu}},\ and\ \bibinfo {author}
		{\bibfnamefont {S.}~\bibnamefont {Chen}},\ }\bibfield  {title} {\bibinfo
		{title} {Interplay of non-hermitian skin effects and anderson localization in
			nonreciprocal quasiperiodic lattices},\ }\href
	{https://doi.org/10.1103/PhysRevB.100.054301} {\bibfield  {journal} {\bibinfo
			{journal} {Phys. Rev. B}\ }\textbf {\bibinfo {volume} {100}},\ \bibinfo
		{pages} {054301} (\bibinfo {year} {2019})}\BibitemShut {NoStop}%
	\bibitem [{\citenamefont {Liu}\ \emph {et~al.}(2020{\natexlab{a}})\citenamefont
		{Liu}, \citenamefont {Jiang}, \citenamefont {Cao},\ and\ \citenamefont
		{Chen}}]{PhysRevB.101.174205}%
	\BibitemOpen
	\bibfield  {author} {\bibinfo {author} {\bibfnamefont {Y.}~\bibnamefont
			{Liu}}, \bibinfo {author} {\bibfnamefont {X.-P.}\ \bibnamefont {Jiang}},
		\bibinfo {author} {\bibfnamefont {J.}~\bibnamefont {Cao}},\ and\ \bibinfo
		{author} {\bibfnamefont {S.}~\bibnamefont {Chen}},\ }\bibfield  {title}
	{\bibinfo {title} {Non-hermitian mobility edges in one-dimensional
			quasicrystals with parity-time symmetry},\ }\href
	{https://doi.org/10.1103/PhysRevB.101.174205} {\bibfield  {journal} {\bibinfo
			{journal} {Phys. Rev. B}\ }\textbf {\bibinfo {volume} {101}},\ \bibinfo
		{pages} {174205} (\bibinfo {year} {2020}{\natexlab{a}})}\BibitemShut
	{NoStop}%
	\bibitem [{\citenamefont {Zeng}\ and\ \citenamefont
		{Xu}(2020)}]{PhysRevResearch.2.033052}%
	\BibitemOpen
	\bibfield  {author} {\bibinfo {author} {\bibfnamefont {Q.-B.}\ \bibnamefont
			{Zeng}}\ and\ \bibinfo {author} {\bibfnamefont {Y.}~\bibnamefont {Xu}},\
	}\bibfield  {title} {\bibinfo {title} {Winding numbers and generalized
			mobility edges in non-hermitian systems},\ }\href
	{https://doi.org/10.1103/PhysRevResearch.2.033052} {\bibfield  {journal}
		{\bibinfo  {journal} {Phys. Rev. Res.}\ }\textbf {\bibinfo {volume} {2}},\
		\bibinfo {pages} {033052} (\bibinfo {year} {2020})}\BibitemShut {NoStop}%
	\bibitem [{\citenamefont {Goblot}\ \emph {et~al.}(2020)\citenamefont {Goblot},
		\citenamefont {Štrkalj}, \citenamefont {Pernet}, \citenamefont {Lado},
		\citenamefont {Dorow}, \citenamefont {Lemaître}, \citenamefont {Le~Gratiet},
		\citenamefont {Harouri}, \citenamefont {Sagnes}, \citenamefont {Ravets},
		\citenamefont {Amo}, \citenamefont {Bloch},\ and\ \citenamefont
		{Zilberberg}}]{Goblot2020}%
	\BibitemOpen
	\bibfield  {author} {\bibinfo {author} {\bibfnamefont {V.}~\bibnamefont
			{Goblot}}, \bibinfo {author} {\bibfnamefont {A.}~\bibnamefont {Štrkalj}},
		\bibinfo {author} {\bibfnamefont {N.}~\bibnamefont {Pernet}}, \bibinfo
		{author} {\bibfnamefont {J.~L.}\ \bibnamefont {Lado}}, \bibinfo {author}
		{\bibfnamefont {C.}~\bibnamefont {Dorow}}, \bibinfo {author} {\bibfnamefont
			{A.}~\bibnamefont {Lemaître}}, \bibinfo {author} {\bibfnamefont
			{L.}~\bibnamefont {Le~Gratiet}}, \bibinfo {author} {\bibfnamefont
			{A.}~\bibnamefont {Harouri}}, \bibinfo {author} {\bibfnamefont
			{I.}~\bibnamefont {Sagnes}}, \bibinfo {author} {\bibfnamefont
			{S.}~\bibnamefont {Ravets}}, \bibinfo {author} {\bibfnamefont
			{A.}~\bibnamefont {Amo}}, \bibinfo {author} {\bibfnamefont {J.}~\bibnamefont
			{Bloch}},\ and\ \bibinfo {author} {\bibfnamefont {O.}~\bibnamefont
			{Zilberberg}},\ }\bibfield  {title} {\bibinfo {title} {Emergence of
			criticality through a cascade of delocalization transitions in quasiperiodic
			chains},\ }\href {https://doi.org/10.1038/s41567-020-0908-7} {\bibfield
		{journal} {\bibinfo  {journal} {Nat. Phys.}\ }\textbf {\bibinfo {volume}
			{16}},\ \bibinfo {pages} {832} (\bibinfo {year} {2020})}\BibitemShut
	{NoStop}%
	\bibitem [{\citenamefont {Longhi}(2021)}]{PhysRevB.103.054203}%
	\BibitemOpen
	\bibfield  {author} {\bibinfo {author} {\bibfnamefont {S.}~\bibnamefont
			{Longhi}},\ }\bibfield  {title} {\bibinfo {title} {Phase transitions in a
			non-hermitian aubry-andr\'e-harper model},\ }\href
	{https://doi.org/10.1103/PhysRevB.103.054203} {\bibfield  {journal} {\bibinfo
			{journal} {Phys. Rev. B}\ }\textbf {\bibinfo {volume} {103}},\ \bibinfo
		{pages} {054203} (\bibinfo {year} {2021})}\BibitemShut {NoStop}%
	\bibitem [{\citenamefont {Wang}\ \emph
		{et~al.}(2022{\natexlab{a}})\citenamefont {Wang}, \citenamefont {Zhang},
		\citenamefont {Li}, \citenamefont {Wu}, \citenamefont {Liu}, \citenamefont
		{Mei}, \citenamefont {Hu}, \citenamefont {Xiao}, \citenamefont {Ma},
		\citenamefont {Chin},\ and\ \citenamefont {Jia}}]{PhysRevLett.129.103401}%
	\BibitemOpen
	\bibfield  {author} {\bibinfo {author} {\bibfnamefont {Y.}~\bibnamefont
			{Wang}}, \bibinfo {author} {\bibfnamefont {J.-H.}\ \bibnamefont {Zhang}},
		\bibinfo {author} {\bibfnamefont {Y.}~\bibnamefont {Li}}, \bibinfo {author}
		{\bibfnamefont {J.}~\bibnamefont {Wu}}, \bibinfo {author} {\bibfnamefont
			{W.}~\bibnamefont {Liu}}, \bibinfo {author} {\bibfnamefont {F.}~\bibnamefont
			{Mei}}, \bibinfo {author} {\bibfnamefont {Y.}~\bibnamefont {Hu}}, \bibinfo
		{author} {\bibfnamefont {L.}~\bibnamefont {Xiao}}, \bibinfo {author}
		{\bibfnamefont {J.}~\bibnamefont {Ma}}, \bibinfo {author} {\bibfnamefont
			{C.}~\bibnamefont {Chin}},\ and\ \bibinfo {author} {\bibfnamefont
			{S.}~\bibnamefont {Jia}},\ }\bibfield  {title} {\bibinfo {title} {Observation
			of interaction-induced mobility edge in an atomic aubry-andr\'e wire},\
	}\href {https://doi.org/10.1103/PhysRevLett.129.103401} {\bibfield  {journal}
		{\bibinfo  {journal} {Phys. Rev. Lett.}\ }\textbf {\bibinfo {volume} {129}},\
		\bibinfo {pages} {103401} (\bibinfo {year} {2022}{\natexlab{a}})}\BibitemShut
	{NoStop}%
	\bibitem [{\citenamefont {Lin}\ \emph {et~al.}(2022)\citenamefont {Lin},
		\citenamefont {Li}, \citenamefont {Xiao}, \citenamefont {Wang}, \citenamefont
		{Yi},\ and\ \citenamefont {Xue}}]{PhysRevLett.129.113601}%
	\BibitemOpen
	\bibfield  {author} {\bibinfo {author} {\bibfnamefont {Q.}~\bibnamefont
			{Lin}}, \bibinfo {author} {\bibfnamefont {T.}~\bibnamefont {Li}}, \bibinfo
		{author} {\bibfnamefont {L.}~\bibnamefont {Xiao}}, \bibinfo {author}
		{\bibfnamefont {K.}~\bibnamefont {Wang}}, \bibinfo {author} {\bibfnamefont
			{W.}~\bibnamefont {Yi}},\ and\ \bibinfo {author} {\bibfnamefont
			{P.}~\bibnamefont {Xue}},\ }\bibfield  {title} {\bibinfo {title} {Topological
			phase transitions and mobility edges in non-hermitian quasicrystals},\ }\href
	{https://doi.org/10.1103/PhysRevLett.129.113601} {\bibfield  {journal}
		{\bibinfo  {journal} {Phys. Rev. Lett.}\ }\textbf {\bibinfo {volume} {129}},\
		\bibinfo {pages} {113601} (\bibinfo {year} {2022})}\BibitemShut {NoStop}%
	\bibitem [{\citenamefont {Zhang}\ and\ \citenamefont
		{Zhang}(2022)}]{PhysRevB.105.174206}%
	\BibitemOpen
	\bibfield  {author} {\bibinfo {author} {\bibfnamefont {Y.-C.}\ \bibnamefont
			{Zhang}}\ and\ \bibinfo {author} {\bibfnamefont {Y.-Y.}\ \bibnamefont
			{Zhang}},\ }\bibfield  {title} {\bibinfo {title} {Lyapunov exponent, mobility
			edges, and critical region in the generalized aubry-andr\'e model with an
			unbounded quasiperiodic potential},\ }\href
	{https://doi.org/10.1103/PhysRevB.105.174206} {\bibfield  {journal} {\bibinfo
			{journal} {Phys. Rev. B}\ }\textbf {\bibinfo {volume} {105}},\ \bibinfo
		{pages} {174206} (\bibinfo {year} {2022})}\BibitemShut {NoStop}%
	\bibitem [{\citenamefont {Weidemann}\ \emph {et~al.}(2022)\citenamefont
		{Weidemann}, \citenamefont {Kremer}, \citenamefont {Longhi},\ and\
		\citenamefont {Szameit}}]{Weidemann2022}%
	\BibitemOpen
	\bibfield  {author} {\bibinfo {author} {\bibfnamefont {S.}~\bibnamefont
			{Weidemann}}, \bibinfo {author} {\bibfnamefont {M.}~\bibnamefont {Kremer}},
		\bibinfo {author} {\bibfnamefont {S.}~\bibnamefont {Longhi}},\ and\ \bibinfo
		{author} {\bibfnamefont {A.}~\bibnamefont {Szameit}},\ }\bibfield  {title}
	{\bibinfo {title} {Topological triple phase transition in non-hermitian
			floquet quasicrystals},\ }\href {https://doi.org/10.1038/s41586-021-04253-0}
	{\bibfield  {journal} {\bibinfo  {journal} {Nature}\ }\textbf {\bibinfo
			{volume} {601}},\ \bibinfo {pages} {354} (\bibinfo {year}
		{2022})}\BibitemShut {NoStop}%
	\bibitem [{\citenamefont {Chang}\ \emph {et~al.}(2025)\citenamefont {Chang},
		\citenamefont {Zhang}, \citenamefont {Lu}, \citenamefont {Yang},
		\citenamefont {Mei}, \citenamefont {Ma}, \citenamefont {Jia},\ and\
		\citenamefont {Jin}}]{PhysRevLett.134.053601}%
	\BibitemOpen
	\bibfield  {author} {\bibinfo {author} {\bibfnamefont {Y.-J.}\ \bibnamefont
			{Chang}}, \bibinfo {author} {\bibfnamefont {J.-H.}\ \bibnamefont {Zhang}},
		\bibinfo {author} {\bibfnamefont {Y.-H.}\ \bibnamefont {Lu}}, \bibinfo
		{author} {\bibfnamefont {Y.-Y.}\ \bibnamefont {Yang}}, \bibinfo {author}
		{\bibfnamefont {F.}~\bibnamefont {Mei}}, \bibinfo {author} {\bibfnamefont
			{J.}~\bibnamefont {Ma}}, \bibinfo {author} {\bibfnamefont {S.}~\bibnamefont
			{Jia}},\ and\ \bibinfo {author} {\bibfnamefont {X.-M.}\ \bibnamefont {Jin}},\
	}\bibfield  {title} {\bibinfo {title} {Observation of photonic mobility edge
			phases},\ }\href {https://doi.org/10.1103/PhysRevLett.134.053601} {\bibfield
		{journal} {\bibinfo  {journal} {Phys. Rev. Lett.}\ }\textbf {\bibinfo
			{volume} {134}},\ \bibinfo {pages} {053601} (\bibinfo {year}
		{2025})}\BibitemShut {NoStop}%
	\bibitem [{\citenamefont {Aubry}\ and\ \citenamefont
		{Andr{\'e}}(1980)}]{aubry1980analyticity}%
	\BibitemOpen
	\bibfield  {author} {\bibinfo {author} {\bibfnamefont {S.}~\bibnamefont
			{Aubry}}\ and\ \bibinfo {author} {\bibfnamefont {G.}~\bibnamefont
			{Andr{\'e}}},\ }\bibfield  {title} {\bibinfo {title} {Analyticity breaking
			and anderson localization in incommensurate lattices},\ }\href@noop {}
	{\bibfield  {journal} {\bibinfo  {journal} {Ann. Israel Phys. Soc}\ }\textbf
		{\bibinfo {volume} {3}},\ \bibinfo {pages} {18} (\bibinfo {year}
		{1980})}\BibitemShut {NoStop}%
	\bibitem [{\citenamefont {Wilkinson}(1984)}]{rspa.1984.0016}%
	\BibitemOpen
	\bibfield  {author} {\bibinfo {author} {\bibfnamefont {M.}~\bibnamefont
			{Wilkinson}},\ }\bibfield  {title} {\bibinfo {title} {Critical properties of
			electron eigenstates in incommensurate systems},\ }\href
	{https://doi.org/10.1098/rspa.1984.0016} {\bibfield  {journal} {\bibinfo
			{journal} {Proceedings of the Royal Society of London. A. Mathematical and
				Physical Sciences}\ }\textbf {\bibinfo {volume} {391}},\ \bibinfo {pages}
		{305} (\bibinfo {year} {1984})}\BibitemShut {NoStop}%
	\bibitem [{\citenamefont {Siebesma}\ and\ \citenamefont
		{Pietronero}(1987)}]{A.P.Siebesma_1987}%
	\BibitemOpen
	\bibfield  {author} {\bibinfo {author} {\bibfnamefont {A.~P.}\ \bibnamefont
			{Siebesma}}\ and\ \bibinfo {author} {\bibfnamefont {L.}~\bibnamefont
			{Pietronero}},\ }\bibfield  {title} {\bibinfo {title} {Multifractal
			properties of wave functions for one-dimensional systems with an
			incommensurate potential},\ }\href
	{https://doi.org/10.1209/0295-5075/4/5/014} {\bibfield  {journal} {\bibinfo
			{journal} {Europhysics Letters}\ }\textbf {\bibinfo {volume} {4}},\ \bibinfo
		{pages} {597} (\bibinfo {year} {1987})}\BibitemShut {NoStop}%
	\bibitem [{\citenamefont {Abe}\ and\ \citenamefont
		{Hiramoto}(1987)}]{PhysRevA.36.5349}%
	\BibitemOpen
	\bibfield  {author} {\bibinfo {author} {\bibfnamefont {S.}~\bibnamefont
			{Abe}}\ and\ \bibinfo {author} {\bibfnamefont {H.}~\bibnamefont {Hiramoto}},\
	}\bibfield  {title} {\bibinfo {title} {Fractal dynamics of electron wave
			packets in one-dimensional quasiperiodic systems},\ }\href
	{https://doi.org/10.1103/PhysRevA.36.5349} {\bibfield  {journal} {\bibinfo
			{journal} {Phys. Rev. A}\ }\textbf {\bibinfo {volume} {36}},\ \bibinfo
		{pages} {5349} (\bibinfo {year} {1987})}\BibitemShut {NoStop}%
	\bibitem [{\citenamefont {Jitomirskaya}(1999)}]{69337d76b53b}%
	\BibitemOpen
	\bibfield  {author} {\bibinfo {author} {\bibfnamefont {S.~Y.}\ \bibnamefont
			{Jitomirskaya}},\ }\bibfield  {title} {\bibinfo {title} {Metal-insulator
			transition for the almost mathieu operator},\ }\href
	{http://www.jstor.org/stable/121066} {\bibfield  {journal} {\bibinfo
			{journal} {Annals of Mathematics}\ }\textbf {\bibinfo {volume} {150}},\
		\bibinfo {pages} {1159} (\bibinfo {year} {1999})}\BibitemShut {NoStop}%
	\bibitem [{\citenamefont {Kohmoto}(1983)}]{PhysRevLett.51.1198}%
	\BibitemOpen
	\bibfield  {author} {\bibinfo {author} {\bibfnamefont {M.}~\bibnamefont
			{Kohmoto}},\ }\bibfield  {title} {\bibinfo {title} {Metal-insulator
			transition and scaling for incommensurate systems},\ }\href
	{https://doi.org/10.1103/PhysRevLett.51.1198} {\bibfield  {journal} {\bibinfo
			{journal} {Phys. Rev. Lett.}\ }\textbf {\bibinfo {volume} {51}},\ \bibinfo
		{pages} {1198} (\bibinfo {year} {1983})}\BibitemShut {NoStop}%
	\bibitem [{\citenamefont {Gopalakrishnan}(2017)}]{PhysRevB.96.054202}%
	\BibitemOpen
	\bibfield  {author} {\bibinfo {author} {\bibfnamefont {S.}~\bibnamefont
			{Gopalakrishnan}},\ }\bibfield  {title} {\bibinfo {title} {Self-dual
			quasiperiodic systems with power-law hopping},\ }\href
	{https://doi.org/10.1103/PhysRevB.96.054202} {\bibfield  {journal} {\bibinfo
			{journal} {Phys. Rev. B}\ }\textbf {\bibinfo {volume} {96}},\ \bibinfo
		{pages} {054202} (\bibinfo {year} {2017})}\BibitemShut {NoStop}%
	\bibitem [{\citenamefont {Deng}\ \emph {et~al.}(2019)\citenamefont {Deng},
		\citenamefont {Ray}, \citenamefont {Sinha}, \citenamefont {Shlyapnikov},\
		and\ \citenamefont {Santos}}]{PhysRevLett.123.025301}%
	\BibitemOpen
	\bibfield  {author} {\bibinfo {author} {\bibfnamefont {X.}~\bibnamefont
			{Deng}}, \bibinfo {author} {\bibfnamefont {S.}~\bibnamefont {Ray}}, \bibinfo
		{author} {\bibfnamefont {S.}~\bibnamefont {Sinha}}, \bibinfo {author}
		{\bibfnamefont {G.~V.}\ \bibnamefont {Shlyapnikov}},\ and\ \bibinfo {author}
		{\bibfnamefont {L.}~\bibnamefont {Santos}},\ }\bibfield  {title} {\bibinfo
		{title} {One-dimensional quasicrystals with power-law hopping},\ }\href
	{https://doi.org/10.1103/PhysRevLett.123.025301} {\bibfield  {journal}
		{\bibinfo  {journal} {Phys. Rev. Lett.}\ }\textbf {\bibinfo {volume} {123}},\
		\bibinfo {pages} {025301} (\bibinfo {year} {2019})}\BibitemShut {NoStop}%
	\bibitem [{\citenamefont {Saha}\ \emph {et~al.}(2019)\citenamefont {Saha},
		\citenamefont {Maiti},\ and\ \citenamefont
		{Purkayastha}}]{PhysRevB.100.174201}%
	\BibitemOpen
	\bibfield  {author} {\bibinfo {author} {\bibfnamefont {M.}~\bibnamefont
			{Saha}}, \bibinfo {author} {\bibfnamefont {S.~K.}\ \bibnamefont {Maiti}},\
		and\ \bibinfo {author} {\bibfnamefont {A.}~\bibnamefont {Purkayastha}},\
	}\bibfield  {title} {\bibinfo {title} {Anomalous transport through
			algebraically localized states in one dimension},\ }\href
	{https://doi.org/10.1103/PhysRevB.100.174201} {\bibfield  {journal} {\bibinfo
			{journal} {Phys. Rev. B}\ }\textbf {\bibinfo {volume} {100}},\ \bibinfo
		{pages} {174201} (\bibinfo {year} {2019})}\BibitemShut {NoStop}%
	\bibitem [{\citenamefont {Fraxanet}\ \emph {et~al.}(2022)\citenamefont
		{Fraxanet}, \citenamefont {Bhattacharya}, \citenamefont {Grass},
		\citenamefont {Lewenstein},\ and\ \citenamefont
		{Dauphin}}]{PhysRevB.106.024204}%
	\BibitemOpen
	\bibfield  {author} {\bibinfo {author} {\bibfnamefont {J.}~\bibnamefont
			{Fraxanet}}, \bibinfo {author} {\bibfnamefont {U.}~\bibnamefont
			{Bhattacharya}}, \bibinfo {author} {\bibfnamefont {T.}~\bibnamefont {Grass}},
		\bibinfo {author} {\bibfnamefont {M.}~\bibnamefont {Lewenstein}},\ and\
		\bibinfo {author} {\bibfnamefont {A.}~\bibnamefont {Dauphin}},\ }\bibfield
	{title} {\bibinfo {title} {Localization and multifractal properties of the
			long-range kitaev chain in the presence of an aubry-andr\'e-harper
			modulation},\ }\href {https://doi.org/10.1103/PhysRevB.106.024204} {\bibfield
		{journal} {\bibinfo  {journal} {Phys. Rev. B}\ }\textbf {\bibinfo {volume}
			{106}},\ \bibinfo {pages} {024204} (\bibinfo {year} {2022})}\BibitemShut
	{NoStop}%
	\bibitem [{\citenamefont {Xia}\ \emph {et~al.}(2022)\citenamefont {Xia},
		\citenamefont {Huang}, \citenamefont {Wang},\ and\ \citenamefont
		{Li}}]{PhysRevB.105.014207}%
	\BibitemOpen
	\bibfield  {author} {\bibinfo {author} {\bibfnamefont {X.}~\bibnamefont
			{Xia}}, \bibinfo {author} {\bibfnamefont {K.}~\bibnamefont {Huang}}, \bibinfo
		{author} {\bibfnamefont {S.}~\bibnamefont {Wang}},\ and\ \bibinfo {author}
		{\bibfnamefont {X.}~\bibnamefont {Li}},\ }\bibfield  {title} {\bibinfo
		{title} {Exact mobility edges in the non-hermitian
			${t}_{1}\text{\ensuremath{-}}{t}_{2}$ model: Theory and possible experimental
			realizations},\ }\href {https://doi.org/10.1103/PhysRevB.105.014207}
	{\bibfield  {journal} {\bibinfo  {journal} {Phys. Rev. B}\ }\textbf {\bibinfo
			{volume} {105}},\ \bibinfo {pages} {014207} (\bibinfo {year}
		{2022})}\BibitemShut {NoStop}%
	\bibitem [{\citenamefont {Wang}\ \emph {et~al.}(2023)\citenamefont {Wang},
		\citenamefont {Zheng}, \citenamefont {Chen}, \citenamefont {Xiao},
		\citenamefont {Jia},\ and\ \citenamefont {Zhang}}]{PhysRevB.107.075128}%
	\BibitemOpen
	\bibfield  {author} {\bibinfo {author} {\bibfnamefont {H.}~\bibnamefont
			{Wang}}, \bibinfo {author} {\bibfnamefont {X.}~\bibnamefont {Zheng}},
		\bibinfo {author} {\bibfnamefont {J.}~\bibnamefont {Chen}}, \bibinfo {author}
		{\bibfnamefont {L.}~\bibnamefont {Xiao}}, \bibinfo {author} {\bibfnamefont
			{S.}~\bibnamefont {Jia}},\ and\ \bibinfo {author} {\bibfnamefont
			{L.}~\bibnamefont {Zhang}},\ }\bibfield  {title} {\bibinfo {title} {Fate of
			the reentrant localization phenomenon in the one-dimensional dimerized
			quasiperiodic chain with long-range hopping},\ }\href
	{https://doi.org/10.1103/PhysRevB.107.075128} {\bibfield  {journal} {\bibinfo
			{journal} {Phys. Rev. B}\ }\textbf {\bibinfo {volume} {107}},\ \bibinfo
		{pages} {075128} (\bibinfo {year} {2023})}\BibitemShut {NoStop}%
	\bibitem [{\citenamefont {Liu}\ and\ \citenamefont
		{Guo}(2018)}]{PhysRevB.98.104201}%
	\BibitemOpen
	\bibfield  {author} {\bibinfo {author} {\bibfnamefont {T.}~\bibnamefont
			{Liu}}\ and\ \bibinfo {author} {\bibfnamefont {H.}~\bibnamefont {Guo}},\
	}\bibfield  {title} {\bibinfo {title} {Mobility edges in off-diagonal
			disordered tight-binding models},\ }\href
	{https://doi.org/10.1103/PhysRevB.98.104201} {\bibfield  {journal} {\bibinfo
			{journal} {Phys. Rev. B}\ }\textbf {\bibinfo {volume} {98}},\ \bibinfo
		{pages} {104201} (\bibinfo {year} {2018})}\BibitemShut {NoStop}%
	\bibitem [{\citenamefont {Griniasty}\ and\ \citenamefont
		{Fishman}(1988)}]{PhysRevLett.60.1334}%
	\BibitemOpen
	\bibfield  {author} {\bibinfo {author} {\bibfnamefont {M.}~\bibnamefont
			{Griniasty}}\ and\ \bibinfo {author} {\bibfnamefont {S.}~\bibnamefont
			{Fishman}},\ }\bibfield  {title} {\bibinfo {title} {Localization by
			pseudorandom potentials in one dimension},\ }\href
	{https://doi.org/10.1103/PhysRevLett.60.1334} {\bibfield  {journal} {\bibinfo
			{journal} {Phys. Rev. Lett.}\ }\textbf {\bibinfo {volume} {60}},\ \bibinfo
		{pages} {1334} (\bibinfo {year} {1988})}\BibitemShut {NoStop}%
	\bibitem [{\citenamefont {Thouless}(1988)}]{PhysRevLett.61.2141}%
	\BibitemOpen
	\bibfield  {author} {\bibinfo {author} {\bibfnamefont {D.~J.}\ \bibnamefont
			{Thouless}},\ }\bibfield  {title} {\bibinfo {title} {Localization by a
			potential with slowly varying period},\ }\href
	{https://doi.org/10.1103/PhysRevLett.61.2141} {\bibfield  {journal} {\bibinfo
			{journal} {Phys. Rev. Lett.}\ }\textbf {\bibinfo {volume} {61}},\ \bibinfo
		{pages} {2141} (\bibinfo {year} {1988})}\BibitemShut {NoStop}%
	\bibitem [{\citenamefont {Das~Sarma}\ \emph {et~al.}(1990)\citenamefont
		{Das~Sarma}, \citenamefont {He},\ and\ \citenamefont
		{Xie}}]{PhysRevB.41.5544}%
	\BibitemOpen
	\bibfield  {author} {\bibinfo {author} {\bibfnamefont {S.}~\bibnamefont
			{Das~Sarma}}, \bibinfo {author} {\bibfnamefont {S.}~\bibnamefont {He}},\ and\
		\bibinfo {author} {\bibfnamefont {X.~C.}\ \bibnamefont {Xie}},\ }\bibfield
	{title} {\bibinfo {title} {Localization, mobility edges, and metal-insulator
			transition in a class of one-dimensional slowly varying deterministic
			potentials},\ }\href {https://doi.org/10.1103/PhysRevB.41.5544} {\bibfield
		{journal} {\bibinfo  {journal} {Phys. Rev. B}\ }\textbf {\bibinfo {volume}
			{41}},\ \bibinfo {pages} {5544} (\bibinfo {year} {1990})}\BibitemShut
	{NoStop}%
	\bibitem [{\citenamefont {Li}\ \emph {et~al.}(2017)\citenamefont {Li},
		\citenamefont {Li},\ and\ \citenamefont {Das~Sarma}}]{PhysRevB.96.085119}%
	\BibitemOpen
	\bibfield  {author} {\bibinfo {author} {\bibfnamefont {X.}~\bibnamefont
			{Li}}, \bibinfo {author} {\bibfnamefont {X.}~\bibnamefont {Li}},\ and\
		\bibinfo {author} {\bibfnamefont {S.}~\bibnamefont {Das~Sarma}},\ }\bibfield
	{title} {\bibinfo {title} {Mobility edges in one-dimensional bichromatic
			incommensurate potentials},\ }\href
	{https://doi.org/10.1103/PhysRevB.96.085119} {\bibfield  {journal} {\bibinfo
			{journal} {Phys. Rev. B}\ }\textbf {\bibinfo {volume} {96}},\ \bibinfo
		{pages} {085119} (\bibinfo {year} {2017})}\BibitemShut {NoStop}%
	\bibitem [{\citenamefont {Yao}\ \emph {et~al.}(2019)\citenamefont {Yao},
		\citenamefont {Khoudli}, \citenamefont {Bresque},\ and\ \citenamefont
		{Sanchez-Palencia}}]{PhysRevLett.123.070405}%
	\BibitemOpen
	\bibfield  {author} {\bibinfo {author} {\bibfnamefont {H.}~\bibnamefont
			{Yao}}, \bibinfo {author} {\bibfnamefont {A.}~\bibnamefont {Khoudli}},
		\bibinfo {author} {\bibfnamefont {L.}~\bibnamefont {Bresque}},\ and\ \bibinfo
		{author} {\bibfnamefont {L.}~\bibnamefont {Sanchez-Palencia}},\ }\bibfield
	{title} {\bibinfo {title} {Critical behavior and fractality in shallow
			one-dimensional quasiperiodic potentials},\ }\href
	{https://doi.org/10.1103/PhysRevLett.123.070405} {\bibfield  {journal}
		{\bibinfo  {journal} {Phys. Rev. Lett.}\ }\textbf {\bibinfo {volume} {123}},\
		\bibinfo {pages} {070405} (\bibinfo {year} {2019})}\BibitemShut {NoStop}%
	\bibitem [{\citenamefont {Hu}\ \emph {et~al.}(2025{\natexlab{a}})\citenamefont
		{Hu}, \citenamefont {Chen}, \citenamefont {Lin}, \citenamefont {Guo},
		\citenamefont {Lin},\ and\ \citenamefont {Gong}}]{2rfb-j778}%
	\BibitemOpen
	\bibfield  {author} {\bibinfo {author} {\bibfnamefont {H.-T.}\ \bibnamefont
			{Hu}}, \bibinfo {author} {\bibfnamefont {Y.}~\bibnamefont {Chen}}, \bibinfo
		{author} {\bibfnamefont {X.}~\bibnamefont {Lin}}, \bibinfo {author}
		{\bibfnamefont {A.-M.}\ \bibnamefont {Guo}}, \bibinfo {author} {\bibfnamefont
			{Z.}~\bibnamefont {Lin}},\ and\ \bibinfo {author} {\bibfnamefont
			{M.}~\bibnamefont {Gong}},\ }\bibfield  {title} {\bibinfo {title} {Exact
			mobility edges in quasiperiodic network models with slowly varying
			potentials},\ }\href {https://doi.org/10.1103/2rfb-j778} {\bibfield
		{journal} {\bibinfo  {journal} {Phys. Rev. B}\ }\textbf {\bibinfo {volume}
			{112}},\ \bibinfo {pages} {054201} (\bibinfo {year}
		{2025}{\natexlab{a}})}\BibitemShut {NoStop}%
	\bibitem [{\citenamefont {Biddle}\ and\ \citenamefont
		{Das~Sarma}(2010)}]{PhysRevLett.104.070601}%
	\BibitemOpen
	\bibfield  {author} {\bibinfo {author} {\bibfnamefont {J.}~\bibnamefont
			{Biddle}}\ and\ \bibinfo {author} {\bibfnamefont {S.}~\bibnamefont
			{Das~Sarma}},\ }\bibfield  {title} {\bibinfo {title} {Predicted mobility
			edges in one-dimensional incommensurate optical lattices: An exactly solvable
			model of anderson localization},\ }\href
	{https://doi.org/10.1103/PhysRevLett.104.070601} {\bibfield  {journal}
		{\bibinfo  {journal} {Phys. Rev. Lett.}\ }\textbf {\bibinfo {volume} {104}},\
		\bibinfo {pages} {070601} (\bibinfo {year} {2010})}\BibitemShut {NoStop}%
	\bibitem [{\citenamefont {Ganeshan}\ \emph {et~al.}(2015)\citenamefont
		{Ganeshan}, \citenamefont {Pixley},\ and\ \citenamefont
		{Das~Sarma}}]{PhysRevLett.114.146601}%
	\BibitemOpen
	\bibfield  {author} {\bibinfo {author} {\bibfnamefont {S.}~\bibnamefont
			{Ganeshan}}, \bibinfo {author} {\bibfnamefont {J.~H.}\ \bibnamefont
			{Pixley}},\ and\ \bibinfo {author} {\bibfnamefont {S.}~\bibnamefont
			{Das~Sarma}},\ }\bibfield  {title} {\bibinfo {title} {Nearest neighbor tight
			binding models with an exact mobility edge in one dimension},\ }\href
	{https://doi.org/10.1103/PhysRevLett.114.146601} {\bibfield  {journal}
		{\bibinfo  {journal} {Phys. Rev. Lett.}\ }\textbf {\bibinfo {volume} {114}},\
		\bibinfo {pages} {146601} (\bibinfo {year} {2015})}\BibitemShut {NoStop}%
	\bibitem [{\citenamefont {Wang}\ \emph
		{et~al.}(2020{\natexlab{a}})\citenamefont {Wang}, \citenamefont {Xia},
		\citenamefont {Zhang}, \citenamefont {Yao}, \citenamefont {Chen},
		\citenamefont {You}, \citenamefont {Zhou},\ and\ \citenamefont
		{Liu}}]{PhysRevLett.125.196604}%
	\BibitemOpen
	\bibfield  {author} {\bibinfo {author} {\bibfnamefont {Y.}~\bibnamefont
			{Wang}}, \bibinfo {author} {\bibfnamefont {X.}~\bibnamefont {Xia}}, \bibinfo
		{author} {\bibfnamefont {L.}~\bibnamefont {Zhang}}, \bibinfo {author}
		{\bibfnamefont {H.}~\bibnamefont {Yao}}, \bibinfo {author} {\bibfnamefont
			{S.}~\bibnamefont {Chen}}, \bibinfo {author} {\bibfnamefont {J.}~\bibnamefont
			{You}}, \bibinfo {author} {\bibfnamefont {Q.}~\bibnamefont {Zhou}},\ and\
		\bibinfo {author} {\bibfnamefont {X.-J.}\ \bibnamefont {Liu}},\ }\bibfield
	{title} {\bibinfo {title} {One-dimensional quasiperiodic mosaic lattice with
			exact mobility edges},\ }\href
	{https://doi.org/10.1103/PhysRevLett.125.196604} {\bibfield  {journal}
		{\bibinfo  {journal} {Phys. Rev. Lett.}\ }\textbf {\bibinfo {volume} {125}},\
		\bibinfo {pages} {196604} (\bibinfo {year} {2020}{\natexlab{a}})}\BibitemShut
	{NoStop}%
	\bibitem [{\citenamefont {Gon\ifmmode~\mbox{\c{c}}\else \c{c}\fi{}alves}\ \emph
		{et~al.}(2023{\natexlab{a}})\citenamefont {Gon\ifmmode~\mbox{\c{c}}\else
			\c{c}\fi{}alves}, \citenamefont {Amorim}, \citenamefont {Castro},\ and\
		\citenamefont {Ribeiro}}]{PhysRevLett.131.186303}%
	\BibitemOpen
	\bibfield  {author} {\bibinfo {author} {\bibfnamefont {M.}~\bibnamefont
			{Gon\ifmmode~\mbox{\c{c}}\else \c{c}\fi{}alves}}, \bibinfo {author}
		{\bibfnamefont {B.}~\bibnamefont {Amorim}}, \bibinfo {author} {\bibfnamefont
			{E.~V.}\ \bibnamefont {Castro}},\ and\ \bibinfo {author} {\bibfnamefont
			{P.}~\bibnamefont {Ribeiro}},\ }\bibfield  {title} {\bibinfo {title}
		{Critical phase dualities in 1d exactly solvable quasiperiodic models},\
	}\href {https://doi.org/10.1103/PhysRevLett.131.186303} {\bibfield  {journal}
		{\bibinfo  {journal} {Phys. Rev. Lett.}\ }\textbf {\bibinfo {volume} {131}},\
		\bibinfo {pages} {186303} (\bibinfo {year} {2023}{\natexlab{a}})}\BibitemShut
	{NoStop}%
	\bibitem [{\citenamefont {Gon\ifmmode~\mbox{\c{c}}\else \c{c}\fi{}alves}\ \emph
		{et~al.}(2023{\natexlab{b}})\citenamefont {Gon\ifmmode~\mbox{\c{c}}\else
			\c{c}\fi{}alves}, \citenamefont {Amorim}, \citenamefont {Castro},\ and\
		\citenamefont {Ribeiro}}]{PhysRevB.108.L100201}%
	\BibitemOpen
	\bibfield  {author} {\bibinfo {author} {\bibfnamefont {M.}~\bibnamefont
			{Gon\ifmmode~\mbox{\c{c}}\else \c{c}\fi{}alves}}, \bibinfo {author}
		{\bibfnamefont {B.}~\bibnamefont {Amorim}}, \bibinfo {author} {\bibfnamefont
			{E.~V.}\ \bibnamefont {Castro}},\ and\ \bibinfo {author} {\bibfnamefont
			{P.}~\bibnamefont {Ribeiro}},\ }\bibfield  {title} {\bibinfo {title}
		{Renormalization group theory of one-dimensional quasiperiodic lattice models
			with commensurate approximants},\ }\href
	{https://doi.org/10.1103/PhysRevB.108.L100201} {\bibfield  {journal}
		{\bibinfo  {journal} {Phys. Rev. B}\ }\textbf {\bibinfo {volume} {108}},\
		\bibinfo {pages} {L100201} (\bibinfo {year}
		{2023}{\natexlab{b}})}\BibitemShut {NoStop}%
	\bibitem [{\citenamefont {Hu}\ \emph {et~al.}(2025{\natexlab{b}})\citenamefont
		{Hu}, \citenamefont {Lin}, \citenamefont {Guo}, \citenamefont {Guo},
		\citenamefont {Lin},\ and\ \citenamefont {Gong}}]{rl1f-ptzq}%
	\BibitemOpen
	\bibfield  {author} {\bibinfo {author} {\bibfnamefont {H.-T.}\ \bibnamefont
			{Hu}}, \bibinfo {author} {\bibfnamefont {X.}~\bibnamefont {Lin}}, \bibinfo
		{author} {\bibfnamefont {A.-M.}\ \bibnamefont {Guo}}, \bibinfo {author}
		{\bibfnamefont {G.}~\bibnamefont {Guo}}, \bibinfo {author} {\bibfnamefont
			{Z.}~\bibnamefont {Lin}},\ and\ \bibinfo {author} {\bibfnamefont
			{M.}~\bibnamefont {Gong}},\ }\bibfield  {title} {\bibinfo {title} {Hidden
			self duality and exact mobility edges in quasiperiodic network models},\
	}\href {https://doi.org/10.1103/rl1f-ptzq} {\bibfield  {journal} {\bibinfo
			{journal} {Phys. Rev. Lett.}\ }\textbf {\bibinfo {volume} {134}},\ \bibinfo
		{pages} {246301} (\bibinfo {year} {2025}{\natexlab{b}})}\BibitemShut
	{NoStop}%
	\bibitem [{\citenamefont {Li}\ \emph {et~al.}(2025)\citenamefont {Li},
		\citenamefont {Zhang}, \citenamefont {Wang}, \citenamefont {Zhang},
		\citenamefont {Zhu},\ and\ \citenamefont {Li}}]{Li2025}%
	\BibitemOpen
	\bibfield  {author} {\bibinfo {author} {\bibfnamefont {S.-Z.}\ \bibnamefont
			{Li}}, \bibinfo {author} {\bibfnamefont {Y.-C.}\ \bibnamefont {Zhang}},
		\bibinfo {author} {\bibfnamefont {Y.}~\bibnamefont {Wang}}, \bibinfo {author}
		{\bibfnamefont {S.}~\bibnamefont {Zhang}}, \bibinfo {author} {\bibfnamefont
			{S.-L.}\ \bibnamefont {Zhu}},\ and\ \bibinfo {author} {\bibfnamefont
			{Z.}~\bibnamefont {Li}},\ }\bibfield  {title} {\bibinfo {title}
		{Multifractal-enriched mobility edges and emergent quantum phases in rydberg
			atomic arrays},\ }\href {https://doi.org/10.1007/s11433-025-2774-2}
	{\bibfield  {journal} {\bibinfo  {journal} {Science China Physics, Mechanics
				\& Astronomy}\ }\textbf {\bibinfo {volume} {69}},\ \bibinfo {pages} {217212}
		(\bibinfo {year} {2025})}\BibitemShut {NoStop}%
	\bibitem [{\citenamefont {Roy}\ \emph {et~al.}(2021)\citenamefont {Roy},
		\citenamefont {Mishra}, \citenamefont {Tanatar},\ and\ \citenamefont
		{Basu}}]{PhysRevLett.126.106803}%
	\BibitemOpen
	\bibfield  {author} {\bibinfo {author} {\bibfnamefont {S.}~\bibnamefont
			{Roy}}, \bibinfo {author} {\bibfnamefont {T.}~\bibnamefont {Mishra}},
		\bibinfo {author} {\bibfnamefont {B.}~\bibnamefont {Tanatar}},\ and\ \bibinfo
		{author} {\bibfnamefont {S.}~\bibnamefont {Basu}},\ }\bibfield  {title}
	{\bibinfo {title} {Reentrant localization transition in a quasiperiodic
			chain},\ }\href {https://doi.org/10.1103/PhysRevLett.126.106803} {\bibfield
		{journal} {\bibinfo  {journal} {Phys. Rev. Lett.}\ }\textbf {\bibinfo
			{volume} {126}},\ \bibinfo {pages} {106803} (\bibinfo {year}
		{2021})}\BibitemShut {NoStop}%
	\bibitem [{\citenamefont {Padhan}\ \emph {et~al.}(2022)\citenamefont {Padhan},
		\citenamefont {Giri}, \citenamefont {Mondal},\ and\ \citenamefont
		{Mishra}}]{PhysRevB.105.L220201}%
	\BibitemOpen
	\bibfield  {author} {\bibinfo {author} {\bibfnamefont {A.}~\bibnamefont
			{Padhan}}, \bibinfo {author} {\bibfnamefont {M.~K.}\ \bibnamefont {Giri}},
		\bibinfo {author} {\bibfnamefont {S.}~\bibnamefont {Mondal}},\ and\ \bibinfo
		{author} {\bibfnamefont {T.}~\bibnamefont {Mishra}},\ }\bibfield  {title}
	{\bibinfo {title} {Emergence of multiple localization transitions in a
			one-dimensional quasiperiodic lattice},\ }\href
	{https://doi.org/10.1103/PhysRevB.105.L220201} {\bibfield  {journal}
		{\bibinfo  {journal} {Phys. Rev. B}\ }\textbf {\bibinfo {volume} {105}},\
		\bibinfo {pages} {L220201} (\bibinfo {year} {2022})}\BibitemShut {NoStop}%
	\bibitem [{\citenamefont {Han}\ and\ \citenamefont
		{Zhou}(2022)}]{PhysRevB.105.054204}%
	\BibitemOpen
	\bibfield  {author} {\bibinfo {author} {\bibfnamefont {W.}~\bibnamefont
			{Han}}\ and\ \bibinfo {author} {\bibfnamefont {L.}~\bibnamefont {Zhou}},\
	}\bibfield  {title} {\bibinfo {title} {Dimerization-induced mobility edges
			and multiple reentrant localization transitions in non-hermitian
			quasicrystals},\ }\href {https://doi.org/10.1103/PhysRevB.105.054204}
	{\bibfield  {journal} {\bibinfo  {journal} {Phys. Rev. B}\ }\textbf {\bibinfo
			{volume} {105}},\ \bibinfo {pages} {054204} (\bibinfo {year}
		{2022})}\BibitemShut {NoStop}%
	\bibitem [{\citenamefont {Qi}\ \emph {et~al.}(2023)\citenamefont {Qi},
		\citenamefont {Cao},\ and\ \citenamefont {Jiang}}]{PhysRevB.107.224201}%
	\BibitemOpen
	\bibfield  {author} {\bibinfo {author} {\bibfnamefont {R.}~\bibnamefont
			{Qi}}, \bibinfo {author} {\bibfnamefont {J.}~\bibnamefont {Cao}},\ and\
		\bibinfo {author} {\bibfnamefont {X.-P.}\ \bibnamefont {Jiang}},\ }\bibfield
	{title} {\bibinfo {title} {Multiple localization transitions and novel
			quantum phases induced by a staggered on-site potential},\ }\href
	{https://doi.org/10.1103/PhysRevB.107.224201} {\bibfield  {journal} {\bibinfo
			{journal} {Phys. Rev. B}\ }\textbf {\bibinfo {volume} {107}},\ \bibinfo
		{pages} {224201} (\bibinfo {year} {2023})}\BibitemShut {NoStop}%
	\bibitem [{\citenamefont {Vaidya}\ \emph {et~al.}(2023)\citenamefont {Vaidya},
		\citenamefont {J\"org}, \citenamefont {Linn}, \citenamefont {Goh},\ and\
		\citenamefont {Rechtsman}}]{PhysRevResearch.5.033170}%
	\BibitemOpen
	\bibfield  {author} {\bibinfo {author} {\bibfnamefont {S.}~\bibnamefont
			{Vaidya}}, \bibinfo {author} {\bibfnamefont {C.}~\bibnamefont {J\"org}},
		\bibinfo {author} {\bibfnamefont {K.}~\bibnamefont {Linn}}, \bibinfo {author}
		{\bibfnamefont {M.}~\bibnamefont {Goh}},\ and\ \bibinfo {author}
		{\bibfnamefont {M.~C.}\ \bibnamefont {Rechtsman}},\ }\bibfield  {title}
	{\bibinfo {title} {Reentrant delocalization transition in one-dimensional
			photonic quasicrystals},\ }\href
	{https://doi.org/10.1103/PhysRevResearch.5.033170} {\bibfield  {journal}
		{\bibinfo  {journal} {Phys. Rev. Res.}\ }\textbf {\bibinfo {volume} {5}},\
		\bibinfo {pages} {033170} (\bibinfo {year} {2023})}\BibitemShut {NoStop}%
	\bibitem [{\citenamefont {Sil}\ \emph {et~al.}(2008)\citenamefont {Sil},
		\citenamefont {Maiti},\ and\ \citenamefont
		{Chakrabarti}}]{PhysRevLett.101.076803}%
	\BibitemOpen
	\bibfield  {author} {\bibinfo {author} {\bibfnamefont {S.}~\bibnamefont
			{Sil}}, \bibinfo {author} {\bibfnamefont {S.~K.}\ \bibnamefont {Maiti}},\
		and\ \bibinfo {author} {\bibfnamefont {A.}~\bibnamefont {Chakrabarti}},\
	}\bibfield  {title} {\bibinfo {title} {Metal-insulator transition in an
			aperiodic ladder network: An exact result},\ }\href
	{https://doi.org/10.1103/PhysRevLett.101.076803} {\bibfield  {journal}
		{\bibinfo  {journal} {Phys. Rev. Lett.}\ }\textbf {\bibinfo {volume} {101}},\
		\bibinfo {pages} {076803} (\bibinfo {year} {2008})}\BibitemShut {NoStop}%
	\bibitem [{\citenamefont {Flach}\ and\ \citenamefont
		{Danieli}(2014)}]{flach2014}%
	\BibitemOpen
	\bibfield  {author} {\bibinfo {author} {\bibfnamefont {S.}~\bibnamefont
			{Flach}}\ and\ \bibinfo {author} {\bibfnamefont {C.}~\bibnamefont
			{Danieli}},\ } {\bibinfo {title}
		{Comment on "metal-insulator transition in an aperiodic ladder network: An
			exact result"}},\ \href {https://arxiv.org/abs/1402.2742} {arXiv:1402.2742 (\bibinfo {year} {2014})} 
	\BibitemShut {NoStop}%
	\bibitem [{\citenamefont {Bodyfelt}\ \emph {et~al.}(2014)\citenamefont
		{Bodyfelt}, \citenamefont {Leykam}, \citenamefont {Danieli}, \citenamefont
		{Yu},\ and\ \citenamefont {Flach}}]{PhysRevLett.113.236403}%
	\BibitemOpen
	\bibfield  {author} {\bibinfo {author} {\bibfnamefont {J.~D.}\ \bibnamefont
			{Bodyfelt}}, \bibinfo {author} {\bibfnamefont {D.}~\bibnamefont {Leykam}},
		\bibinfo {author} {\bibfnamefont {C.}~\bibnamefont {Danieli}}, \bibinfo
		{author} {\bibfnamefont {X.}~\bibnamefont {Yu}},\ and\ \bibinfo {author}
		{\bibfnamefont {S.}~\bibnamefont {Flach}},\ }\bibfield  {title} {\bibinfo
		{title} {Flatbands under correlated perturbations},\ }\href
	{https://doi.org/10.1103/PhysRevLett.113.236403} {\bibfield  {journal}
		{\bibinfo  {journal} {Phys. Rev. Lett.}\ }\textbf {\bibinfo {volume} {113}},\
		\bibinfo {pages} {236403} (\bibinfo {year} {2014})}\BibitemShut {NoStop}%
	\bibitem [{\citenamefont {Bordia}\ \emph {et~al.}(2016)\citenamefont {Bordia},
		\citenamefont {L\"uschen}, \citenamefont {Hodgman}, \citenamefont
		{Schreiber}, \citenamefont {Bloch},\ and\ \citenamefont
		{Schneider}}]{PhysRevLett.116.140401}%
	\BibitemOpen
	\bibfield  {author} {\bibinfo {author} {\bibfnamefont {P.}~\bibnamefont
			{Bordia}}, \bibinfo {author} {\bibfnamefont {H.~P.}\ \bibnamefont
			{L\"uschen}}, \bibinfo {author} {\bibfnamefont {S.~S.}\ \bibnamefont
			{Hodgman}}, \bibinfo {author} {\bibfnamefont {M.}~\bibnamefont {Schreiber}},
		\bibinfo {author} {\bibfnamefont {I.}~\bibnamefont {Bloch}},\ and\ \bibinfo
		{author} {\bibfnamefont {U.}~\bibnamefont {Schneider}},\ }\bibfield  {title}
	{\bibinfo {title} {Coupling identical one-dimensional many-body localized
			systems},\ }\href {https://doi.org/10.1103/PhysRevLett.116.140401} {\bibfield
		{journal} {\bibinfo  {journal} {Phys. Rev. Lett.}\ }\textbf {\bibinfo
			{volume} {116}},\ \bibinfo {pages} {140401} (\bibinfo {year}
		{2016})}\BibitemShut {NoStop}%
	\bibitem [{\citenamefont {An}\ \emph {et~al.}(2018)\citenamefont {An},
		\citenamefont {Meier},\ and\ \citenamefont {Gadway}}]{PhysRevX.8.031045}%
	\BibitemOpen
	\bibfield  {author} {\bibinfo {author} {\bibfnamefont {F.~A.}\ \bibnamefont
			{An}}, \bibinfo {author} {\bibfnamefont {E.~J.}\ \bibnamefont {Meier}},\ and\
		\bibinfo {author} {\bibfnamefont {B.}~\bibnamefont {Gadway}},\ }\bibfield
	{title} {\bibinfo {title} {Engineering a flux-dependent mobility edge in
			disordered zigzag chains},\ }\href
	{https://doi.org/10.1103/PhysRevX.8.031045} {\bibfield  {journal} {\bibinfo
			{journal} {Phys. Rev. X}\ }\textbf {\bibinfo {volume} {8}},\ \bibinfo {pages}
		{031045} (\bibinfo {year} {2018})}\BibitemShut {NoStop}%
	\bibitem [{\citenamefont {Rossignolo}\ and\ \citenamefont
		{Dell'Anna}(2019)}]{PhysRevB.99.054211}%
	\BibitemOpen
	\bibfield  {author} {\bibinfo {author} {\bibfnamefont {M.}~\bibnamefont
			{Rossignolo}}\ and\ \bibinfo {author} {\bibfnamefont {L.}~\bibnamefont
			{Dell'Anna}},\ }\bibfield  {title} {\bibinfo {title} {Localization
			transitions and mobility edges in coupled aubry-andr\'e chains},\ }\href
	{https://doi.org/10.1103/PhysRevB.99.054211} {\bibfield  {journal} {\bibinfo
			{journal} {Phys. Rev. B}\ }\textbf {\bibinfo {volume} {99}},\ \bibinfo
		{pages} {054211} (\bibinfo {year} {2019})}\BibitemShut {NoStop}%
	\bibitem [{\citenamefont {Lopes~dos Santos}\ \emph {et~al.}(2007)\citenamefont
		{Lopes~dos Santos}, \citenamefont {Peres},\ and\ \citenamefont
		{Castro~Neto}}]{PhysRevLett.99.256802}%
	\BibitemOpen
	\bibfield  {author} {\bibinfo {author} {\bibfnamefont {J.~M.~B.}\
			\bibnamefont {Lopes~dos Santos}}, \bibinfo {author} {\bibfnamefont
			{N.~M.~R.}\ \bibnamefont {Peres}},\ and\ \bibinfo {author} {\bibfnamefont
			{A.~H.}\ \bibnamefont {Castro~Neto}},\ }\bibfield  {title} {\bibinfo {title}
		{Graphene bilayer with a twist: Electronic structure},\ }\href
	{https://doi.org/10.1103/PhysRevLett.99.256802} {\bibfield  {journal}
		{\bibinfo  {journal} {Phys. Rev. Lett.}\ }\textbf {\bibinfo {volume} {99}},\
		\bibinfo {pages} {256802} (\bibinfo {year} {2007})}\BibitemShut {NoStop}%
	\bibitem [{\citenamefont {Trambly~de Laissardière}\ \emph
		{et~al.}(2010)\citenamefont {Trambly~de Laissardière}, \citenamefont
		{Mayou},\ and\ \citenamefont {Magaud}}]{Trambly2010}%
	\BibitemOpen
	\bibfield  {author} {\bibinfo {author} {\bibfnamefont {G.}~\bibnamefont
			{Trambly~de Laissardière}}, \bibinfo {author} {\bibfnamefont
			{D.}~\bibnamefont {Mayou}},\ and\ \bibinfo {author} {\bibfnamefont
			{L.}~\bibnamefont {Magaud}},\ }\bibfield  {title} {\bibinfo {title}
		{Localization of dirac electrons in rotated graphene bilayers},\ }\href
	{https://doi.org/10.1021/nl902948m} {\bibfield  {journal} {\bibinfo
			{journal} {Nano Lett.}\ }\textbf {\bibinfo {volume} {10}},\ \bibinfo
		{pages} {804} (\bibinfo {year} {2010})}\BibitemShut {NoStop}%
	\bibitem [{\citenamefont {Bistritzer}\ and\ \citenamefont
		{MacDonald}(2011)}]{pnas.1108174108}%
	\BibitemOpen
	\bibfield  {author} {\bibinfo {author} {\bibfnamefont {R.}~\bibnamefont
			{Bistritzer}}\ and\ \bibinfo {author} {\bibfnamefont {A.~H.}\ \bibnamefont
			{MacDonald}},\ }\bibfield  {title} {\bibinfo {title} {Moiré bands in twisted
			double-layer graphene},\ }\href {https://doi.org/10.1073/pnas.1108174108}
	{\bibfield  {journal} {\bibinfo  {journal} {PNAS USA}\ }\textbf {\bibinfo {volume} {108}},\ \bibinfo {pages}
		{12233} (\bibinfo {year} {2011})}\BibitemShut {NoStop}%
	\bibitem [{}]{Suppl}%
	\BibitemOpen
	\bibfield  {title}	{\bibinfo {title} {See Supplementary Materials for more details on
			(a) phase shift representation, (b) decoupled equations for $\phi=0$ and $\phi=\pi$, (c) phase diagram of quasi-periodic ladders with $\phi=\pi$, 
			(d) proof of chiral symmetry for even lattice constant, 
			(e) renormalized quasi-periodic potential strength $2\tilde{t}$, 
			(f) band structure evolution and 
			(g) phase diagram of quasi-periodic ladders for $\kappa=4$. Including Ref.~\cite{PhysRevLett.126.106803,PhysRevB.105.L220201}}\ }\BibitemShut 
	{NoStop}
	\bibitem [{\citenamefont {Wang}\ \emph {et~al.}(2021)\citenamefont {Wang},
		\citenamefont {Xia}, \citenamefont {Wang}, \citenamefont {Zheng},\ and\
		\citenamefont {Liu}}]{PhysRevB.103.174205}%
	\BibitemOpen
	\bibfield  {author} {\bibinfo {author} {\bibfnamefont {Y.}~\bibnamefont
			{Wang}}, \bibinfo {author} {\bibfnamefont {X.}~\bibnamefont {Xia}}, \bibinfo
		{author} {\bibfnamefont {Y.}~\bibnamefont {Wang}}, \bibinfo {author}
		{\bibfnamefont {Z.}~\bibnamefont {Zheng}},\ and\ \bibinfo {author}
		{\bibfnamefont {X.-J.}\ \bibnamefont {Liu}},\ }\bibfield  {title} {\bibinfo
		{title} {Duality between two generalized aubry-andr\'e models with exact
			mobility edges},\ }\href {https://doi.org/10.1103/PhysRevB.103.174205}
	{\bibfield  {journal} {\bibinfo  {journal} {Phys. Rev. B}\ }\textbf {\bibinfo
			{volume} {103}},\ \bibinfo {pages} {174205} (\bibinfo {year}
		{2021})}\BibitemShut {NoStop}%
	\bibitem [{\citenamefont {Gonçalves}\ \emph {et~al.}(2022)\citenamefont
		{Gonçalves}, \citenamefont {Amorim}, \citenamefont {Castro},\ and\
		\citenamefont {Ribeiro}}]{SciPostPhys.13.3.046}%
	\BibitemOpen
	\bibfield  {author} {\bibinfo {author} {\bibfnamefont {M.}~\bibnamefont
			{Gonçalves}}, \bibinfo {author} {\bibfnamefont {B.}~\bibnamefont {Amorim}},
		\bibinfo {author} {\bibfnamefont {E.~V.}\ \bibnamefont {Castro}},\ and\
		\bibinfo {author} {\bibfnamefont {P.}~\bibnamefont {Ribeiro}},\ }\bibfield
	{title} {\bibinfo {title} {{Hidden dualities in 1D quasiperiodic lattice
				models}},\ }\href {https://doi.org/10.21468/SciPostPhys.13.3.046} {\bibfield
		{journal} {\bibinfo  {journal} {SciPost Phys.}\ }\textbf {\bibinfo {volume}
			{13}},\ \bibinfo {pages} {046} (\bibinfo {year} {2022})}\BibitemShut
	{NoStop}%
	\bibitem [{\citenamefont {Gadway}(2015)}]{PhysRevA.92.043606}%
	\BibitemOpen
	\bibfield  {author} {\bibinfo {author} {\bibfnamefont {B.}~\bibnamefont
			{Gadway}},\ }\bibfield  {title} {\bibinfo {title} {Atom-optics approach to
			studying transport phenomena},\ }\href
	{https://doi.org/10.1103/PhysRevA.92.043606} {\bibfield  {journal} {\bibinfo
			{journal} {Phys. Rev. A}\ }\textbf {\bibinfo {volume} {92}},\ \bibinfo
		{pages} {043606} (\bibinfo {year} {2015})}\BibitemShut {NoStop}%
	\bibitem [{\citenamefont {An}\ \emph {et~al.}(2017)\citenamefont {An},
		\citenamefont {Meier},\ and\ \citenamefont {Gadway}}]{sciadv.1602685}%
	\BibitemOpen
	\bibfield  {author} {\bibinfo {author} {\bibfnamefont {F.~A.}\ \bibnamefont
			{An}}, \bibinfo {author} {\bibfnamefont {E.~J.}\ \bibnamefont {Meier}},\ and\
		\bibinfo {author} {\bibfnamefont {B.}~\bibnamefont {Gadway}},\ }\bibfield
	{title} {\bibinfo {title} {Direct observation of chiral currents and magnetic
			reflection in atomic flux lattices},\ }\href
	{https://doi.org/10.1126/sciadv.1602685} {\bibfield  {journal} {\bibinfo
			{journal} {Sci. Adv.}\ }\textbf {\bibinfo {volume} {3}},\ \bibinfo
		{pages} {e1602685} (\bibinfo {year} {2017})}\BibitemShut {NoStop}%
	\bibitem [{\citenamefont {Liang}\ \emph {et~al.}(2024)\citenamefont {Liang},
		\citenamefont {Dong}, \citenamefont {Pan}, \citenamefont {Wang},
		\citenamefont {Li}, \citenamefont {Yang}, \citenamefont {Yi},\ and\
		\citenamefont {Yan}}]{liang2024chiral}%
	\BibitemOpen
	\bibfield  {author} {\bibinfo {author} {\bibfnamefont {Q.}~\bibnamefont
			{Liang}}, \bibinfo {author} {\bibfnamefont {Z.}~\bibnamefont {Dong}},
		\bibinfo {author} {\bibfnamefont {J.-S.}\ \bibnamefont {Pan}}, \bibinfo
		{author} {\bibfnamefont {H.}~\bibnamefont {Wang}}, \bibinfo {author}
		{\bibfnamefont {H.}~\bibnamefont {Li}}, \bibinfo {author} {\bibfnamefont
			{Z.}~\bibnamefont {Yang}}, \bibinfo {author} {\bibfnamefont {W.}~\bibnamefont
			{Yi}},\ and\ \bibinfo {author} {\bibfnamefont {B.}~\bibnamefont {Yan}},\
	}\bibfield  {title} {\bibinfo {title} {Chiral dynamics of ultracold atoms
			under a tunable su(2) synthetic gauge field},\ } {\bibfield
		{journal} {\bibinfo  {journal} {Nat. Phys.}\ }\textbf {\bibinfo {volume}
			{20}},\ \bibinfo {pages} {1738} (\bibinfo {year} {2024})}\BibitemShut
	{NoStop}%
	\bibitem [{\citenamefont {Cai}\ \emph {et~al.}(2019)\citenamefont {Cai},
		\citenamefont {Liu}, \citenamefont {Wu}, \citenamefont {He}, \citenamefont
		{Zhu}, \citenamefont {Zhang},\ and\ \citenamefont
		{Wang}}]{PhysRevLett.122.023601}%
	\BibitemOpen
	\bibfield  {author} {\bibinfo {author} {\bibfnamefont {H.}~\bibnamefont
			{Cai}}, \bibinfo {author} {\bibfnamefont {J.}~\bibnamefont {Liu}}, \bibinfo
		{author} {\bibfnamefont {J.}~\bibnamefont {Wu}}, \bibinfo {author}
		{\bibfnamefont {Y.}~\bibnamefont {He}}, \bibinfo {author} {\bibfnamefont
			{S.-Y.}\ \bibnamefont {Zhu}}, \bibinfo {author} {\bibfnamefont {J.-X.}\
			\bibnamefont {Zhang}},\ and\ \bibinfo {author} {\bibfnamefont {D.-W.}\
			\bibnamefont {Wang}},\ }\bibfield  {title} {\bibinfo {title} {Experimental
			observation of momentum-space chiral edge currents in room-temperature
			atoms},\ }\href {https://doi.org/10.1103/PhysRevLett.122.023601} {\bibfield
		{journal} {\bibinfo  {journal} {Phys. Rev. Lett.}\ }\textbf {\bibinfo
			{volume} {122}},\ \bibinfo {pages} {023601} (\bibinfo {year}
		{2019})}\BibitemShut {NoStop}%
	\bibitem [{\citenamefont {Hainaut}\ \emph {et~al.}(2018)\citenamefont
		{Hainaut}, \citenamefont {Manai}, \citenamefont {Cl{\'e}ment}, \citenamefont
		{Garreau}, \citenamefont {Szriftgiser}, \citenamefont {Lemari{\'e}},
		\citenamefont {Cherroret}, \citenamefont {Delande},\ and\ \citenamefont
		{Chicireanu}}]{hainaut2018controlling}%
	\BibitemOpen
	\bibfield  {author} {\bibinfo {author} {\bibfnamefont {C.}~\bibnamefont
			{Hainaut}}, \bibinfo {author} {\bibfnamefont {I.}~\bibnamefont {Manai}},
		\bibinfo {author} {\bibfnamefont {J.-F.}\ \bibnamefont {Cl{\'e}ment}},
		\bibinfo {author} {\bibfnamefont {J.~C.}\ \bibnamefont {Garreau}}, \bibinfo
		{author} {\bibfnamefont {P.}~\bibnamefont {Szriftgiser}}, \bibinfo {author}
		{\bibfnamefont {G.}~\bibnamefont {Lemari{\'e}}}, \bibinfo {author}
		{\bibfnamefont {N.}~\bibnamefont {Cherroret}}, \bibinfo {author}
		{\bibfnamefont {D.}~\bibnamefont {Delande}},\ and\ \bibinfo {author}
		{\bibfnamefont {R.}~\bibnamefont {Chicireanu}},\ }\bibfield  {title}
	{\bibinfo {title} {Controlling symmetry and localization with an artificial
			gauge field in a disordered quantum system},\ }\href@noop {} {\bibfield
		{journal} {\bibinfo  {journal} {Nature communications}\ }\textbf {\bibinfo
			{volume} {9}},\ \bibinfo {pages} {1382} (\bibinfo {year} {2018})}\BibitemShut
	{NoStop}%
	\bibitem [{\citenamefont {Li}\ and\ \citenamefont
		{Das~Sarma}(2020)}]{PhysRevB.101.064203}%
	\BibitemOpen
	\bibfield  {author} {\bibinfo {author} {\bibfnamefont {X.}~\bibnamefont
			{Li}}\ and\ \bibinfo {author} {\bibfnamefont {S.}~\bibnamefont {Das~Sarma}},\
	}\bibfield  {title} {\bibinfo {title} {Mobility edge and intermediate phase
			in one-dimensional incommensurate lattice potentials},\ }\href
	{https://doi.org/10.1103/PhysRevB.101.064203} {\bibfield  {journal} {\bibinfo
			{journal} {Phys. Rev. B}\ }\textbf {\bibinfo {volume} {101}},\ \bibinfo
		{pages} {064203} (\bibinfo {year} {2020})}\BibitemShut {NoStop}%
	\bibitem [{\citenamefont {Hofstadter}(1976)}]{PhysRevB.14.2239}%
	\BibitemOpen
	\bibfield  {author} {\bibinfo {author} {\bibfnamefont {D.~R.}\ \bibnamefont
			{Hofstadter}},\ }\bibfield  {title} {\bibinfo {title} {Energy levels and wave
			functions of bloch electrons in rational and irrational magnetic fields},\
	}\href {https://doi.org/10.1103/PhysRevB.14.2239} {\bibfield  {journal}
		{\bibinfo  {journal} {Phys. Rev. B}\ }\textbf {\bibinfo {volume} {14}},\
		\bibinfo {pages} {2239} (\bibinfo {year} {1976})}\BibitemShut {NoStop}%
	\bibitem [{\citenamefont {Huang}\ \emph {et~al.}(2020)\citenamefont {Huang},
		\citenamefont {Hosur},\ and\ \citenamefont {Pal}}]{PhysRevB.102.155429}%
	\BibitemOpen
	\bibfield  {author} {\bibinfo {author} {\bibfnamefont {Y.}~\bibnamefont
			{Huang}}, \bibinfo {author} {\bibfnamefont {P.}~\bibnamefont {Hosur}},\ and\
		\bibinfo {author} {\bibfnamefont {H.~K.}\ \bibnamefont {Pal}},\ }\bibfield
	{title} {\bibinfo {title} {Quasi-flat-band physics in a two-leg ladder model
			and its relation to magic-angle twisted bilayer graphene},\ }\href
	{https://doi.org/10.1103/PhysRevB.102.155429} {\bibfield  {journal} {\bibinfo
			{journal} {Phys. Rev. B}\ }\textbf {\bibinfo {volume} {102}},\ \bibinfo
		{pages} {155429} (\bibinfo {year} {2020})}\BibitemShut {NoStop}%
	\bibitem [{\citenamefont {Wiersma}\ \emph {et~al.}(1997)\citenamefont
		{Wiersma}, \citenamefont {Bartolini}, \citenamefont {Lagendijk},\ and\
		\citenamefont {Righini}}]{Wiersma}%
	\BibitemOpen
	\bibfield  {author} {\bibinfo {author} {\bibfnamefont {D.~S.}\ \bibnamefont
			{Wiersma}}, \bibinfo {author} {\bibfnamefont {P.}~\bibnamefont {Bartolini}},
		\bibinfo {author} {\bibfnamefont {A.}~\bibnamefont {Lagendijk}},\ and\
		\bibinfo {author} {\bibfnamefont {R.}~\bibnamefont {Righini}},\ }\bibfield
	{title} {\bibinfo {title} {Localization of light in a disordered medium},\
	}\href {https://doi.org/10.1038/37757} {\bibfield  {journal} {\bibinfo
			{journal} {Nature}\ }\textbf {\bibinfo {volume} {390}},\ \bibinfo {pages}
		{671} (\bibinfo {year} {1997})}\BibitemShut {NoStop}%
	\bibitem [{\citenamefont {Tanese}\ \emph {et~al.}(2014)\citenamefont {Tanese},
		\citenamefont {Gurevich}, \citenamefont {Baboux}, \citenamefont {Jacqmin},
		\citenamefont {Lema\^{\i}tre}, \citenamefont {Galopin}, \citenamefont
		{Sagnes}, \citenamefont {Amo}, \citenamefont {Bloch},\ and\ \citenamefont
		{Akkermans}}]{PhysRevLett.112.146404}%
	\BibitemOpen
	\bibfield  {author} {\bibinfo {author} {\bibfnamefont {D.}~\bibnamefont
			{Tanese}}, \bibinfo {author} {\bibfnamefont {E.}~\bibnamefont {Gurevich}},
		\bibinfo {author} {\bibfnamefont {F.}~\bibnamefont {Baboux}}, \bibinfo
		{author} {\bibfnamefont {T.}~\bibnamefont {Jacqmin}}, \bibinfo {author}
		{\bibfnamefont {A.}~\bibnamefont {Lema\^{\i}tre}}, \bibinfo {author}
		{\bibfnamefont {E.}~\bibnamefont {Galopin}}, \bibinfo {author} {\bibfnamefont
			{I.}~\bibnamefont {Sagnes}}, \bibinfo {author} {\bibfnamefont
			{A.}~\bibnamefont {Amo}}, \bibinfo {author} {\bibfnamefont {J.}~\bibnamefont
			{Bloch}},\ and\ \bibinfo {author} {\bibfnamefont {E.}~\bibnamefont
			{Akkermans}},\ }\bibfield  {title} {\bibinfo {title} {Fractal energy spectrum
			of a polariton gas in a fibonacci quasiperiodic potential},\ }\href
	{https://doi.org/10.1103/PhysRevLett.112.146404} {\bibfield  {journal}
		{\bibinfo  {journal} {Phys. Rev. Lett.}\ }\textbf {\bibinfo {volume} {112}},\
		\bibinfo {pages} {146404} (\bibinfo {year} {2014})}\BibitemShut {NoStop}%
	\bibitem [{\citenamefont {Wang}\ \emph
		{et~al.}(2020{\natexlab{b}})\citenamefont {Wang}, \citenamefont {Zheng},
		\citenamefont {Chen}, \citenamefont {Huang}, \citenamefont {Kartashov},
		\citenamefont {Torner}, \citenamefont {Konotop},\ and\ \citenamefont
		{Ye}}]{Wang2020}%
	\BibitemOpen
	\bibfield  {author} {\bibinfo {author} {\bibfnamefont {P.}~\bibnamefont
			{Wang}}, \bibinfo {author} {\bibfnamefont {Y.}~\bibnamefont {Zheng}},
		\bibinfo {author} {\bibfnamefont {X.}~\bibnamefont {Chen}}, \bibinfo {author}
		{\bibfnamefont {C.}~\bibnamefont {Huang}}, \bibinfo {author} {\bibfnamefont
			{Y.~V.}\ \bibnamefont {Kartashov}}, \bibinfo {author} {\bibfnamefont
			{L.}~\bibnamefont {Torner}}, \bibinfo {author} {\bibfnamefont {V.~V.}\
			\bibnamefont {Konotop}},\ and\ \bibinfo {author} {\bibfnamefont
			{F.}~\bibnamefont {Ye}},\ }\bibfield  {title} {\bibinfo {title} {Localization
			and delocalization of light in photonic moiré lattices},\ }\href
	{https://doi.org/10.1038/s41586-019-1851-6} {\bibfield  {journal} {\bibinfo
			{journal} {Nature}\ }\textbf {\bibinfo {volume} {577}},\ \bibinfo {pages}
		{42} (\bibinfo {year} {2020}{\natexlab{b}})}\BibitemShut {NoStop}%
	\bibitem [{\citenamefont {Liu}\ \emph {et~al.}(2020{\natexlab{b}})\citenamefont
		{Liu}, \citenamefont {Yang}, \citenamefont {Ren}, \citenamefont {Xue},
		\citenamefont {Lin}, \citenamefont {Hu}, \citenamefont {Sun}, \citenamefont
		{Peng}, \citenamefont {Zhou}, \citenamefont {Chong},\ and\ \citenamefont
		{Zhang}}]{PhysRevLett.125.133603}%
	\BibitemOpen
	\bibfield  {author} {\bibinfo {author} {\bibfnamefont {G.-G.}\ \bibnamefont
			{Liu}}, \bibinfo {author} {\bibfnamefont {Y.}~\bibnamefont {Yang}}, \bibinfo
		{author} {\bibfnamefont {X.}~\bibnamefont {Ren}}, \bibinfo {author}
		{\bibfnamefont {H.}~\bibnamefont {Xue}}, \bibinfo {author} {\bibfnamefont
			{X.}~\bibnamefont {Lin}}, \bibinfo {author} {\bibfnamefont {Y.-H.}\
			\bibnamefont {Hu}}, \bibinfo {author} {\bibfnamefont {H.-x.}\ \bibnamefont
			{Sun}}, \bibinfo {author} {\bibfnamefont {B.}~\bibnamefont {Peng}}, \bibinfo
		{author} {\bibfnamefont {P.}~\bibnamefont {Zhou}}, \bibinfo {author}
		{\bibfnamefont {Y.}~\bibnamefont {Chong}},\ and\ \bibinfo {author}
		{\bibfnamefont {B.}~\bibnamefont {Zhang}},\ }\bibfield  {title} {\bibinfo
		{title} {Topological anderson insulator in disordered photonic crystals},\
	}\href {https://doi.org/10.1103/PhysRevLett.125.133603} {\bibfield  {journal}
		{\bibinfo  {journal} {Phys. Rev. Lett.}\ }\textbf {\bibinfo {volume} {125}},\
		\bibinfo {pages} {133603} (\bibinfo {year} {2020}{\natexlab{b}})}\BibitemShut
	{NoStop}%
	\bibitem [{\citenamefont {Xiao}\ \emph {et~al.}(2021)\citenamefont {Xiao},
		\citenamefont {Xie}, \citenamefont {Dong}, \citenamefont {Chen},
		\citenamefont {Yi},\ and\ \citenamefont {Yan}}]{XIAO20212175}%
	\BibitemOpen
	\bibfield  {author} {\bibinfo {author} {\bibfnamefont {T.}~\bibnamefont
			{Xiao}}, \bibinfo {author} {\bibfnamefont {D.}~\bibnamefont {Xie}}, \bibinfo
		{author} {\bibfnamefont {Z.}~\bibnamefont {Dong}}, \bibinfo {author}
		{\bibfnamefont {T.}~\bibnamefont {Chen}}, \bibinfo {author} {\bibfnamefont
			{W.}~\bibnamefont {Yi}},\ and\ \bibinfo {author} {\bibfnamefont
			{B.}~\bibnamefont {Yan}},\ }\bibfield  {title} {\bibinfo {title} {Observation
			of topological phase with critical localization in a quasi-periodic
			lattice},\ }\href
	{https://doi.org/https://doi.org/10.1016/j.scib.2021.07.025} {\bibfield
		{journal} {\bibinfo  {journal} {Sci. Bull.}\ }\textbf {\bibinfo
			{volume} {66}},\ \bibinfo {pages} {2175} (\bibinfo {year}
		{2021})}\BibitemShut {NoStop}%
	\bibitem [{\citenamefont {Chen}\ \emph {et~al.}(2021)\citenamefont {Chen},
		\citenamefont {Zhu}, \citenamefont {Tan}, \citenamefont {Wang},\ and\
		\citenamefont {Ma}}]{PhysRevX.11.011016}%
	\BibitemOpen
	\bibfield  {author} {\bibinfo {author} {\bibfnamefont {Z.-G.}\ \bibnamefont
			{Chen}}, \bibinfo {author} {\bibfnamefont {W.}~\bibnamefont {Zhu}}, \bibinfo
		{author} {\bibfnamefont {Y.}~\bibnamefont {Tan}}, \bibinfo {author}
		{\bibfnamefont {L.}~\bibnamefont {Wang}},\ and\ \bibinfo {author}
		{\bibfnamefont {G.}~\bibnamefont {Ma}},\ }\bibfield  {title} {\bibinfo
		{title} {Acoustic realization of a four-dimensional higher-order chern
			insulator and boundary-modes engineering},\ }\href
	{https://doi.org/10.1103/PhysRevX.11.011016} {\bibfield  {journal} {\bibinfo
			{journal} {Phys. Rev. X}\ }\textbf {\bibinfo {volume} {11}},\ \bibinfo
		{pages} {011016} (\bibinfo {year} {2021})}\BibitemShut {NoStop}%
	\bibitem [{\citenamefont {Wang}\ \emph
		{et~al.}(2022{\natexlab{b}})\citenamefont {Wang}, \citenamefont {Chen},
		\citenamefont {Gong}, \citenamefont {Yu}, \citenamefont {Chen}, \citenamefont
		{Tian}, \citenamefont {Ren},\ and\ \citenamefont
		{Sun}}]{PhysRevLett.129.173601}%
	\BibitemOpen
	\bibfield  {author} {\bibinfo {author} {\bibfnamefont {L.-C.}\ \bibnamefont
			{Wang}}, \bibinfo {author} {\bibfnamefont {Y.}~\bibnamefont {Chen}}, \bibinfo
		{author} {\bibfnamefont {M.}~\bibnamefont {Gong}}, \bibinfo {author}
		{\bibfnamefont {F.}~\bibnamefont {Yu}}, \bibinfo {author} {\bibfnamefont
			{Q.-D.}\ \bibnamefont {Chen}}, \bibinfo {author} {\bibfnamefont {Z.-N.}\
			\bibnamefont {Tian}}, \bibinfo {author} {\bibfnamefont {X.-F.}\ \bibnamefont
			{Ren}},\ and\ \bibinfo {author} {\bibfnamefont {H.-B.}\ \bibnamefont {Sun}},\
	}\bibfield  {title} {\bibinfo {title} {Edge state, localization length, and
			critical exponent from survival probability in topological waveguides},\
	}\href {https://doi.org/10.1103/PhysRevLett.129.173601} {\bibfield  {journal}
		{\bibinfo  {journal} {Phys. Rev. Lett.}\ }\textbf {\bibinfo {volume} {129}},\
		\bibinfo {pages} {173601} (\bibinfo {year} {2022}{\natexlab{b}})}\BibitemShut
	{NoStop}%
	\bibitem [{\citenamefont {Ren}\ \emph {et~al.}(2024)\citenamefont {Ren},
		\citenamefont {Yu}, \citenamefont {Wu}, \citenamefont {Qi}, \citenamefont
		{Wang}, \citenamefont {Yao}, \citenamefont {Ren}, \citenamefont {Guo},
		\citenamefont {Jiang}, \citenamefont {Chen}, \citenamefont {Liu},
		\citenamefont {Chen},\ and\ \citenamefont {Sun}}]{PhysRevLett.132.066602}%
	\BibitemOpen
	\bibfield  {author} {\bibinfo {author} {\bibfnamefont {M.}~\bibnamefont
			{Ren}}, \bibinfo {author} {\bibfnamefont {Y.}~\bibnamefont {Yu}}, \bibinfo
		{author} {\bibfnamefont {B.}~\bibnamefont {Wu}}, \bibinfo {author}
		{\bibfnamefont {X.}~\bibnamefont {Qi}}, \bibinfo {author} {\bibfnamefont
			{Y.}~\bibnamefont {Wang}}, \bibinfo {author} {\bibfnamefont {X.}~\bibnamefont
			{Yao}}, \bibinfo {author} {\bibfnamefont {J.}~\bibnamefont {Ren}}, \bibinfo
		{author} {\bibfnamefont {Z.}~\bibnamefont {Guo}}, \bibinfo {author}
		{\bibfnamefont {H.}~\bibnamefont {Jiang}}, \bibinfo {author} {\bibfnamefont
			{H.}~\bibnamefont {Chen}}, \bibinfo {author} {\bibfnamefont {X.-J.}\
			\bibnamefont {Liu}}, \bibinfo {author} {\bibfnamefont {Z.}~\bibnamefont
			{Chen}},\ and\ \bibinfo {author} {\bibfnamefont {Y.}~\bibnamefont {Sun}},\
	}\bibfield  {title} {\bibinfo {title} {Realization of gapped and ungapped
			photonic topological anderson insulators},\ }\href
	{https://doi.org/10.1103/PhysRevLett.132.066602} {\bibfield  {journal}
		{\bibinfo  {journal} {Phys. Rev. Lett.}\ }\textbf {\bibinfo {volume} {132}},\
		\bibinfo {pages} {066602} (\bibinfo {year} {2024})}\BibitemShut {NoStop}%
	\bibitem [{\citenamefont {Hou}\ \emph {et~al.}(2026)\citenamefont {Hou},
		\citenamefont {Wu}, \citenamefont {Wang}, \citenamefont {Zhu}, \citenamefont
		{Yan},\ and\ \citenamefont {Yang}}]{Hou2026}%
	\BibitemOpen
	\bibfield  {author} {\bibinfo {author} {\bibfnamefont {X.}~\bibnamefont
			{Hou}}, \bibinfo {author} {\bibfnamefont {Z.}~\bibnamefont {Wu}}, \bibinfo
		{author} {\bibfnamefont {F.}~\bibnamefont {Wang}}, \bibinfo {author}
		{\bibfnamefont {S.}~\bibnamefont {Zhu}}, \bibinfo {author} {\bibfnamefont
			{B.}~\bibnamefont {Yan}},\ and\ \bibinfo {author} {\bibfnamefont
			{Z.}~\bibnamefont {Yang}},\ }\bibfield  {title} {\bibinfo {title} {Quantum
			boomerang effect of light},\ }\bibfield  {journal} {\bibinfo  {journal}
		{Nat. Commun.}\ }\href	{10.1038/s41467-026-68293-8} \bibinfo {\bf 17},
\bibinfo {pages} {1579} (\bibinfo {year} {2026}) \BibitemShut {NoStop}%
\end{thebibliography}
\end{document}